\documentclass[preprint,12pt]{elsarticle}
\usepackage{hyperref}

\usepackage{amssymb}
\usepackage{pdfpages}

\journal{Communications in Nonlinear Science and Numerical Simulation}

\begin{document}

\begin{frontmatter}



\title{On the influence of input triggering on the dynamics of the Jansen-Rit oscillators network}




\author[mymainaddress]{Sheida Kazemi}

\author[mymainaddress]{Yousef Jamali\corref{mycorrespondingauthor}}
\cortext[mycorrespondingauthor]{Y.jamali@modares.ac.ir}

\address[mymainaddress]{Biomathematics Laboratory, Department of Applied Mathematics, School of Mathematical Sciences, Tarbiat Modares University, Tehran, Iran}

\begin{abstract}
We investigate the dynamical properties of a network of coupled neural oscillators with identical Jansen-Rit masses. The connections between nodes follow the regular Watts-Strogatz topology. Each node receives deterministic external input plus internal input based on the output signal from the neighbors.  This paper aims to change these two inputs and analyze the generated results. First, we attempt to analyze the model using the mean-field approximation, i.e., identical inputs for all nodes. Then, this assumption is relaxed. A more detailed analysis of these two states is discussed. We use the Pearson correlation coefficient to measure the amount of synchronization. As a result of the mean-field approach, we find that despite observed changes in behavior, there is no phase transition. Surprisingly, both the first (discontinuous) and second (continuous) phase transition occurs by relaxing the mean-field assumption.
We also demonstrate how changes in input can result in pathological oscillations similar to those observed in epilepsy. Results show that coupled Jansen-Rit masses emerge a variety of behaviors influenced by various external and internal inputs. Moreover, our findings indicate that delta waves can be generated by altering these inputs in a network of Jansen-Rit neural mass models, which has not been previously observed in a single Jansen-Rit neural mass model analysis. \\
Overall, a wide range of behavioral patterns, including epilepsy, healthy behavior, and the transition between synchrony, and asynchrony are examined comprehensively in this paper. Moreover, this work highlights the putative contribution of external and internal inputs in studying the phase transition and synchronization of neural mass models. 
\end{abstract}



\section*{Highlights}
\begin{itemize}
\item The analysis of oscillatory behavior in a coupled nonlinear system is considered
\item The change of coupling term and external input results in a transition to different dynamical states
\item The first (discontinuous) and second (continuous) phase transitions result from relaxing the mean-field assumption
\item Delta waves can be generated by altering inputs in a network of Jansen-Rit neural mass models, which has not been previously observed in a single Jansen-Rit neural mass analysis
\end{itemize}

\begin{keyword}
Brain dynamics, Neural mass model, Synchronization, Bifurcation



\end{keyword}

\end{frontmatter}


\section{Introduction}
Complex networks are ubiquitous in nature, and studies of their nonlinear dynamics and synchronization provide insight into physical, chemical, and biological phenomena \citep{bar1998dynamics, strogatz2001exploring, boccaletti2006complex, nicholls2001neuron}. Modeling real neural systems with complex mathematical networks is beneficial \citep{dayan2001theoretical, deutsch2007mathematical}. Brain networks can be represented as graphs with nodes (neurons or brain regions) connected by edges (physical connections) \citep{beim2008lectures}. Coupled systems play a substantial role in science and engineering \citep{boccaletti2018synchronization}. Research in the nonlinear dynamic analysis is increasingly focused on complex networks with interacting elements \citep{wunderling2021modelling, hebbink2020analysis}. 
The empirical measurement of individual structural connectivity in many research, such as \citep{spiegler2010bifurcation}, has been used to examine networks of coupled phase oscillators. 
Regular, random, small-world, and scale-free networks are four important network topologies that have been studied extensively. As structural links between cortical neurons demonstrate a small-world topology, several studies have investigated small-world networks of interconnected units \citep{yu2017stochastic, odor2019critical, budzinski2019synchronization}.

A significant amount of research has been dedicated to synchronizing coupled systems \citep{castanedo2018synchronization, budzinski2019synchronization, liu2020synchronization}. Synchronization phenomena significantly affect how the brain functions, both normally \citep{mizuhara2007human} and abnormally \citep{uhlhaas2006neural}.  Multiple brain disorders, including Parkinson's disease \citep{he2015contribution}, autism \citep{dinstein2011disrupted}, and epilepsy \citep{liu2017transition}, may result in pathological brain synchronization. Nonlinear dynamics is one of the most fundamental phenomena in phase synchronization \citep{pikovsky2002synchronization}. 

The neural mass models have been around for a long time. These models describe a local population of interacting neurons. Also, they are capable of modeling diverse phenomena and generating a wide variety of dynamic behaviors. Coupling a collection of the neural mass model into a mesoscale circuit can provide a link between micro and macro scales \citep{gosak2018network}. It is essential to model a network of neural masses in order to examine brain function and determine how connected units behave differently. 
A comprehensive understanding of neural populations has recently been gained at the macro-and mesoscales. Neural mass models are a valuable tool for explaining the changes in neuronal dynamics between different behaviors, as they correspond directly to the large-scale dynamics measured by EEG \citep{coombes2005waves, deco2008dynamic}.
Some models that investigated the collective behavior of neurons included Lopez da Silva \citep{da1974model, da1989interdependence}, Jansen and Rit \citep{jansen1995electroencephalogram, jansen1993neurophysiologically}, Wendling \citep{wendling2000relevance, wendling2002epileptic}, Wilson-Cowan \citep{wilson1972excitatory, wilson1973mathematical}, Freeman \citep{freeman1987simulation, chang1996parameter}, and Wong-Wang \citep{wong2006recurrent, deco2013resting}.

Since 1988, analyses and numerical studies of the Jansen-Rit model have been conducted \citep{jansen1995electroencephalogram, jansen1993neurophysiologically, grimbert2006bifurcation, kuhlmann2016neural, Ableidinger2017stochastic, kazemi2022phase}. This study investigates the dynamics of the Jansen-Rit model, which has emerged as a way to extend simultaneous electrical activities simulation, specifically the alpha rhythm, in neural masses. Additionally, Jansen and Rit demonstrated that their model simulated the evoked potentials \citep{jansen1993neurophysiologically}. The main disadvantage of this model is that it fails to produce different rhythms, especially at the onset activity of epilepsy. 
The use of different inputs in a single Jansen-Rit neural mass model \citep{spiegler2010bifurcation, grimbert2006bifurcation} and two coupled Jansen-Rit neural mass models \citep{ahmadizadeh2018bifurcation} can induce various types of activity, such as alpha-like activity and seizure-like activity. The transitions between regimes in dynamics are mathematically modeled using the dynamical systems theory. Consequently, phase transitions are manifested as so-called bifurcations of system variables.

In this paper, we use mesoscale modeling to show the brain as a network of Jansen-Rit coupled oscillators, where the network connection architecture consists of a regular Watts-Strogatz topology with 50 nodes.
Each node receives an external input plus an internal input based on the output signal from the neighbors.

The aim of this paper is twofold. As a first step, we use a mean-field approximation to assume that all nodes have the same input. This implies that all nodes are fully synchronized. The bifurcation diagram presents various types of bifurcation. Second, this mean-field assumption will then be relaxed, and different types of synchronization will be observed as a function of internal and external inputs. In order to quantify synchronization, we used four measures included Pearson Cross-Correlation, Phase Locking Value, Kuramoto Order Parameter, and Phase Locking Index \citep{bastos2016tutorial}. These all yielded almost the same quality results. To simplify, only one was chosen (Pearson CrossCorrelation). In the mean-field approach, we find that there is no phase transition despite observed behavior changes. It is surprising to find that both the first (discontinuous) and second (continuous) phase transitions result from relaxing the mean-field assumption.\\
Following this, we investigate the network behavior in frequency space. Our results show a wide range of rhythms, including delta, theta, and alpha.
We investigate how inputs changes can result in pathological oscillations similar to those observed in seizure.
According to our research, delta waves can be generated by altering the inputs in a network of Jansen-Rit neural mass models, which has not been previously observed in a single Jansen-Rit neural mass analysis.  This phenomenon is an emergent property in a complex system. This states that every isolated node represents an alpha rhythm that changes when these units interact with each other in a complex system.

\section{Methods}\label{method}
\subsection*{Model formulation}
\begin{figure}[!ht]
\centering
\includegraphics[width=8cm, height=10cm]{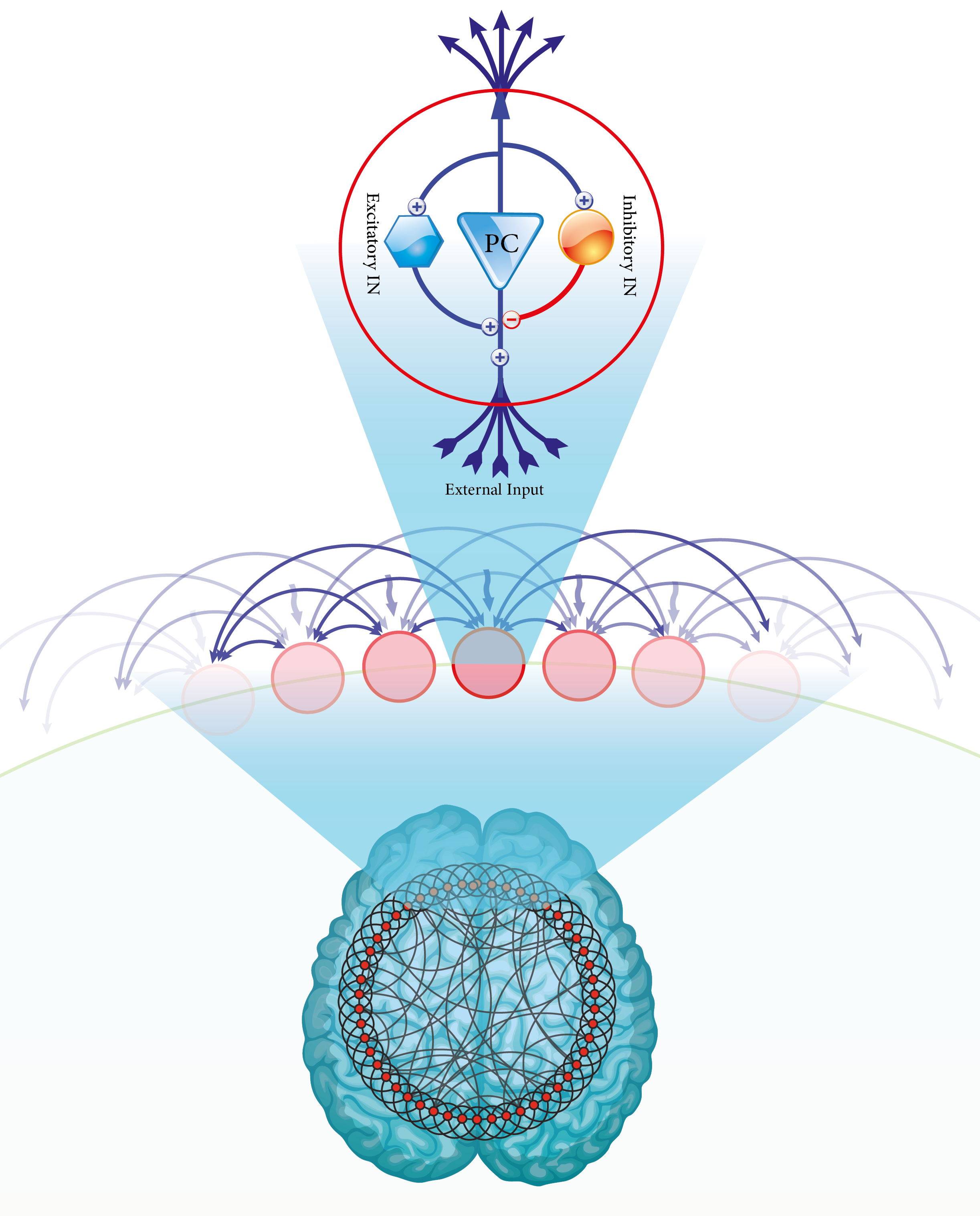}
\caption{Schematic representation of Jansen-Rit dynamics in a Watts-Strogatz network. Each node contains the main population (pyramidal cell (PC)) that interacts with two populations: excitatory interneuron (blue hexagonal on the left side) and inhibitory interneuron (orange circle on the right side).}
\label{J1}
\end{figure}

In this section, the mathematical representation of the Jansen-Rit neural mass model is explained. It describes a cortical area.
This model consists of three parts: pyramidal neurons, excitatory neurons, and inhibitory neurons. Pyramidal neurons are the main population that receives inputs from three sources: excitatory and inhibitory interneuron feedback within the same column, and external input from other columns. A schematic illustration of each mass is shown in figure \ref{J1}.
 
The dynamics of a single unit is described by \citep{grimbert2006bifurcation}
\begin{eqnarray*} 
\begin{array}{lr} 
\dot{x}_{0} (t) = x_{3} (t)\\ 
\dot{x}_{3} (t) = A a S(x_{1} (t) - x_{2} (t)) -2a x_{3} (t) -a^{2}x_{0} (t)\\ 
\dot{x}_{1} (t) = x_{4} (t)\\ 
\dot{x}_{4} (t) = A a \left\lbrace f_{0}(t)+C_{2} S[C_{1} x_{0} (t)]\right\rbrace  -2ax_{4}(t) -a^{2} x_{1}(t)\\ 
\dot{x}_{2} (t) = x_{5} (t)\\ 
\dot{x}_{5} (t) = B b C_{4} S(C_{3}x_{0}) -2b x_{5} (t) -b^{2}x_{2} (t) 
\end{array} 
\end{eqnarray*}

The activities of pyramidal, excitatory, and inhibitory ensembles are ($ x_{0} $;$ x_{3} $), ($ x_{1} $;$ x_{4} $) and ($ x_{2} $;$ x_{5} $), respectively.\\
The sigmoidal function S transforms the average membrane potential of neurons to the mean firing rate of action potentials and is defined as:
\begin{eqnarray*}
S(v) = \frac{v_{max}}{1 + e^{r(v_{0}-v)}}
\end{eqnarray*}
A biological interpretation of the parameter values used to solve equation (1) is given in Table 1 \citep{jansen1995electroencephalogram}.\\
\begin{table} 
\caption{Parameters in the Jansen-Rit model are obtained experimentally.} 
\label{Table} 
\begin{tabular}{ccc} 
\hline 
Parameters & Interpretation & Value 
\\ 
\hline 
$ A $ & Excitatory PSP amplitude &  3.25 mV 
\\ 
\hline \ 
$ B $ & inhibitory PSP amplitude &  22 mV 
\\ 
\hline 
$ 1/a $ &  Time constant of excitatory PSP  &  $ a = 100 s^{-1} $ 
\\ 
\hline 
$ 1/b $ &  Time constant of inhibitory PSP  &  $ b = 50 s^{-1} $ 
\\ 
\hline 
$ C1 $, $ C2 $ & Average numbers of synapses between excitatory populations &  $ 1 * C $, $ 0.8 * C $ 
\\ 
\hline 
$ C3 $, $ C4 $ & Average numbers of synapses between inhibitory populations &  $ 0.25 * C $ 
\\ 
\hline 
$ C $ & Average numbers of synapses between neural populations &  135 
\\ 
\hline 
$ v_{max} $ & Maximum firing rate &  5 Hz 
\\ 
\hline 
$ v0 $ & Potential at half of maximum firing rate &  6 mV 
\\ 
\hline 
$ r $ &  Slope of sigmoid function at $ v0 $ &  $ 0.56 mV^{-1} $ 
\\ 
\hline 
\end{tabular} 
\end{table} 

The value of $ f_{0} $ describes an external input that originated from an external source. Based on the interactions between two populations of interneurons, one excitatory and one inhibitory, the model output is ($ x_{1} $ - $ x_{2} $), which corresponds to the membrane potential of pyramidal cells \citep{kandel2000principles}. Also, $ C_{i} $ = $ C \alpha_{i} $, for i =1,\ldots, 4.\\
In order to construct a network of interactive neural mass models, we assume that the pyramidal cell receives input from both its neighbors and the external input from other sources. In a network with $ N $ nodes, the Jansen-Rit dynamical equations are given as follows (i=1,\ldots, N) \citep{forrester2020role}:
\begin{eqnarray}
\begin{array}{lr} 
\dot{x}_{0i} (t) = x_{3i} (t)\\ 
\dot{x}_{3i} (t) = A a S(x_{1i} (t) - x_{2i} (t)) -2a x_{3i} (t) -a^{2}x_{0i} (t)\\ 
\dot{x}_{1i} (t) = x_{4i} (t)\\ 
\dot{x}_{4i} (t) = A a \lbrace f_{0i}(t)+ \alpha \sum_{j=1}^{N} M_{ij} S(x_{1j} (t) - x_{2j} (t)) + C_{2} S[C_{1} x_{0i} (t)]\rbrace -2ax_{4i}(t) -a^{2} x_{1i}(t)\\ 
\dot{x}_{2i} (t) = x_{5i} (t)\\ 
\dot{x}_{5i} (t) = B b C_{4} S(C_{3}x_{0i}) -2b x_{5i} (t) -b^{2}x_{2i} (t) 
\end{array} \label{MJ}
\end{eqnarray}

$ M $ is the adjacency matrix of the network describing how nodes interact and $ \alpha $ is the coupling term between nodes.\\
The parameters of all units are assumed to be the same, and changes in the network behavior result from varying external inputs and coupling coefficient between units.

\subsection{Synchronization quantifier}
Multiple criteria can be used to measure neural interactions, each with its own advantages and disadvantages \citep{bastos2016tutorial}. We quantify the synchronization between neural mass models using the concept of Pearson cross-correlation (Pcc). This quantifier identifies the linear relationship between each pair of random variables. In this measure of correlation, functional connections between individual neurons are determined \citep{lachaux1999measuring}.  The Pcc coefficient between two time-series $ x $ and $ y $ is defined as follows: 
\begin{eqnarray*}
r = \frac{\sum_{i=1} ^{N}(x_{i} - \langle x \rangle) (y_{i} - \langle y \rangle)}{\sqrt{\sum_{i=1} ^{N}(x_{i} - \langle x \rangle)^2}  \sqrt{\sum_{i=1} ^{N}(y_{i} - \langle y \rangle)^2}} 
\end{eqnarray*} 
Where $ <x> $ is the mean of time series $ x $ and N represents the length of the signal. This coefficient varies between -1 and 1. An amount of 1 (-1) indicates a full linear positive (negative) correlation between two series, whereas 0 signifies no correlation between two series. This definition constructs a symmetric matrix that elements are one in the main diagonal. 

\subsection{Bifurcations}
A bifurcation refers to a qualitative change in dynamics caused by changes in the parameters of a dynamical system. Bifurcations have codimensions corresponding to the minimum number of parameters that are required for a system to display the type of bifurcation \citep{kuznetsov2013elements}. \\
Two general classifications of local codimension 1 bifurcations are saddle-node bifurcations and Hopf bifurcations that distinguished by the sign of Jacobian eigenvalues. 
The saddle-node (SN) bifurcation is identified by a zero eigenvalue. An SN bifurcation is often undesirable in biological systems, as it can lead the system to depart from its previous operation point, such that functionality is lost \citep{novick1957enzyme, griffith1968mathematics}. 
Hopf bifurcations are determined by a Jacobian with a pair of conjugate complex eigenvalues with a zero real part. Their occurrence corresponds to the appearance and disappearance of limit cycles. Therefore, oscillatory dynamics models display this kind of bifurcation. We can further distinguish between supercritical and subcritical Hopf bifurcations by observing on which side of the bifurcation there is a limit cycle.
When a stable steady-state transition to an unstable one results in the emergence of a stable limit cycle, this is known as a supercritical Hopf bifurcation that involves sustained oscillations. Subcritical Hopf bifurcations occur when an unstable cycle coexists with a stable equilibrium on one side of the bifurcation. When an equilibrium becomes unstable, the limit cycles disappear.

A bifurcation of codimension 2 requires tuning two parameters to specific values. Codimension 2 bifurcations in a model are not only of interest theoretically but can also provide information about system dynamics such as bistability and chaos. The following is a brief overview of the most common bifurcations in codimension 2.

\subsubsection*{Generalized Hopf bifurcation:} The Generalized Hopf (GH) bifurcations \citep{bautin1984behavior, shil2001methods} occur at the transition between supercritical and subcritical Hopf bifurcations.
\subsubsection*{Bogdanov-Takens bifurcation:}The Bogdanov-Takens (BT) bifurcation \citep{takens1974singularities, carrillo2010analysis} exhibits two zero real eigenvalues of the Jacobian matrix. An BT bifurcation occurs when a saddle-node bifurcation collides with a Hopf bifurcation.

\subsubsection*{Cusp bifurcation:} A cusp bifurcation point (CP) \citep{guardia2010analytical} is a specific type of SN bifurcation that can be used to identify bistable regions.\\

It is noteworthy that on a two-parameter bifurcation diagram, codimension-1 bifurcations appear as one-dimensional lines \citep{guckenheimer2013nonlinear}.

\subsection{Equilibria}
By setting the left-hand side of (\ref{MJ}) to zero, the network equilibria (called fixed-point) is determined, thus starting the bifurcation analysis. Let $ x_{i} $ = ($ x_{0i} $,\ldots,$ x_{5i})^{T} $ for i=1, \ldots, N. The system has the following form:
\begin{equation}
‌\dot{x}_{i}=F(x_{i};f_{0},\alpha)
\end{equation}
Where F is the smooth map from $ R^{6} $ to $ R^{6} $ according to (\ref{MJ}) and $ f_{0} $ and $ \alpha $ are the inputs from external sources and the coupling term between nodes, respectively. Writing $ \dot{x}_{i}=0 $ for each i=1,\ldots, N. This gives us the system of equations (i=1,\ldots, N):

\begin{eqnarray}
\label{equilib} 
\begin{array}{lr} 
\dot{x}_{3i} (t) = 0\\ 
\dot{x}_{4i} (t) = 0\\
\dot{x}_{5i} (t) = 0\\ 
\dot{x}_{0i} (t) = \frac{A}{a} S(x_{1i} (t) - x_{2i} (t))\\  
\dot{x}_{1i} (t) = \frac{A}{a} \lbrace f_{0i}(t)+ \alpha \sum_{j=1}^{N} M_{ij} S(x_{1j} (t) - x_{2j} (t)) + C_{2} S[C_{1} x_{0i} (t)]\rbrace \\ 
\dot{x}_{2i} (t) = \frac{B}{b} C_{4} S(C_{3}x_{0i}) 
\end{array} 
\end{eqnarray}
According to the EEG signal, the variable ($ x_{1}-x_{2} $) reflects the firing rate of pyramidal cells. In the cortex's superficial layer, which accounts for the majority of the EEG, the apical dendrites of pyramidal neurons deliver their postsynaptic potentials. So, we define $ \mathsf{x}_{i}:=x_{1i} - x_{2i} $ which leads to the following implicit equations:
\begin{equation}\label{eqib_2}
\mathsf{x}_{i}(t) = \frac{A}{a} \left[ f_{0} + \alpha \sum_{j=1}^{N} M_{ij} S(\mathsf{x}_{i}(t)) + C_{2} S[C_{1} \frac{A}{a}S(\mathsf{x}_{i}(t))] \right] - \frac{B}{b} C_{4} S(C_{3} \frac{A}{a} S(\mathsf{x}_{i}(t)))
\end{equation}
for i=1, \ldots, N.\\
It is important to note that we assume all local parameters are identical, thus allowing network behavior only to change as inputs vary.

\subsection{Simulation}
In this simulation, a regular Watts-Strogatz network
was considered with 50 phase oscillators as nodes. Nodes are connected to eight neighbors, four on either side. We investigated the Jansen-Rit dynamics in all nodes. Runge–Kutta algorithm was chosen to integrate the system using a time step of $ 10^{4} $ s. Each simulation was 200 seconds long. On each coupling coefficient, each simulation twelve times. First, we used high-intensity noise to prevent bias in special conditions (up to 2.5s). Then, the noise was interrupted, and after some fluctuations, the equilibrium state was reached. By discarding the first 100 seconds of simulation from further analysis, we were confident that the system had reached to its equilibrium. Every node received an input from another part of the brain as an external input. Different aspects of network behavior are explored when the coupling coefficient between nodes and external inputs varies widely. Moreover, the initial conditions were changed in each repetition.

\section{Results and discussion}
In order to solve (\ref{eqib_2}), two approaches are available. First, we assume identical inputs for nodes by using the mean-field approximation ($ \mathsf{x}_{1} = \ldots, \mathsf{x}_{N} = \mathsf{x} $). In the second case, it is assumed that nodes do not create identical outputs. These two states are examined separately.
\subsection{Analysis of network behavior based on mean-field approach:}
\begin{figure}[!ht]
\centering
\includegraphics[width=10cm, height=8cm]{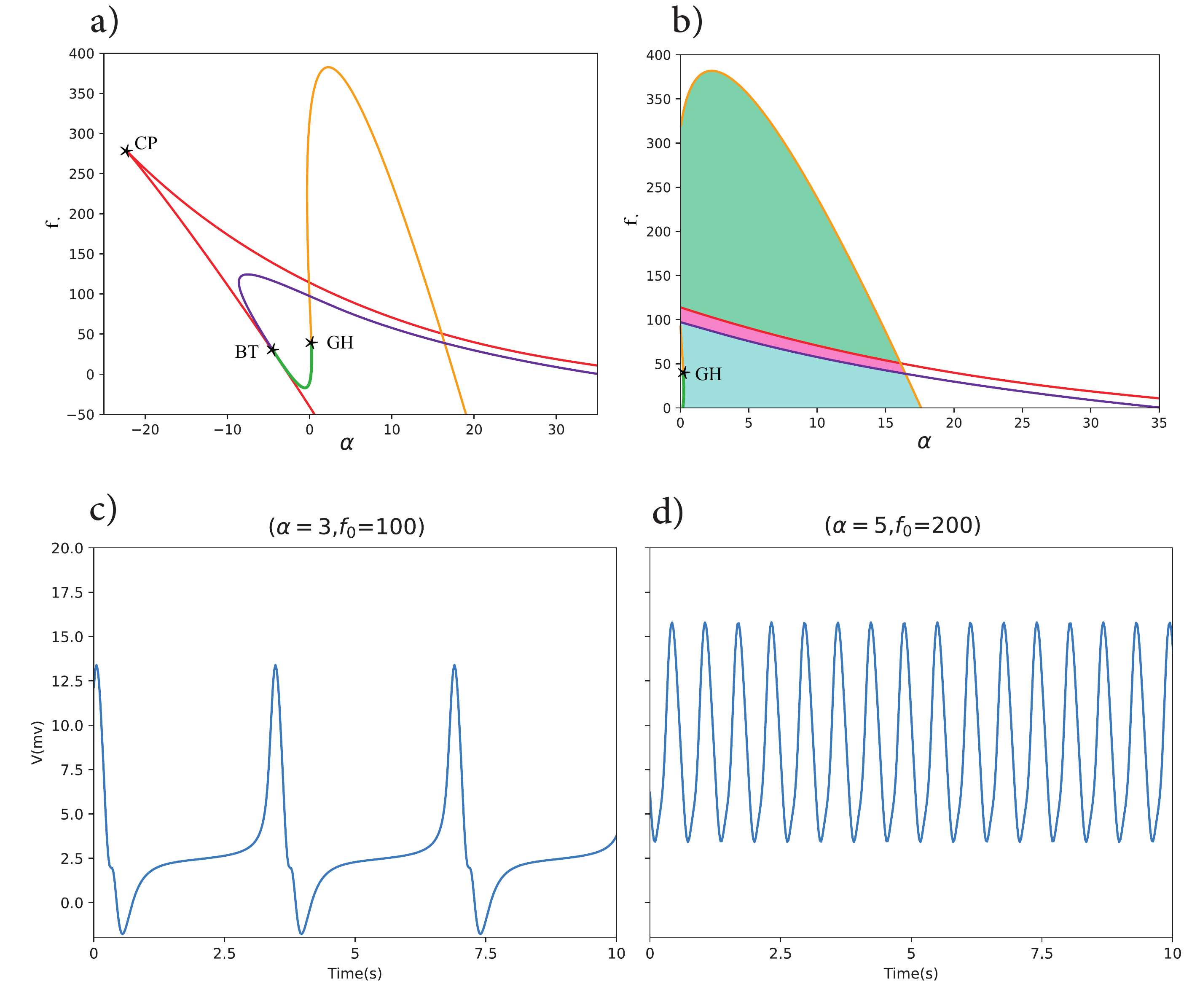}
\caption{\textbf{(a)} Codimension 2 bifurcations of Jansen-Rit model in a network in the ($ \alpha $, $ f_{0} $)-plane. Each color and label enjoy different dynamical properties. Codimension 1 bifurcation curves are shown in red: saddle-node, green: subcritical Hopf, orange: supercritical Hopf, purple: saddle homoclinic. Codimension 2 bifurcation points: CP: Cusp, BT: Bogdanov-Takens, GH: Generalized Hopf. \textbf{(b)} shows the specific area ($ \alpha $, $ f_{0} $) of \textbf{(a)} that is positive. The light blue color region indicates the steady-state activity. The pink color displays spiking behavior with a large amplitude (panel \textbf{(c)}). An oscillatory pattern can be seen in the green zone, which correlates with alpha rhythms (panel \textbf{(d)}).}
\label{bif_1}
\end{figure}

In this case, equation (\ref{eqib_2}) converts to the following equation:
\begin{equation}
\mathsf{x}(t) = \frac{A}{a} \left[ f_{0} + \alpha \sum_{j=1}^{N} M_{ij} S(\mathsf{x}(t)) + C_{2} S[C_{1} \frac{A}{a}S(\mathsf{x}_{i}(t))] \right] - \frac{B}{b} C_{4} S(C_{3} \frac{A}{a} S(\mathsf{x}(t))).
\end{equation}
Using the MatCont package in Matlab, we investigate bifurcations numerically. Indeed, Matcont is a MATLAB software package for numerically exploring dynamical systems in particular bifurcation studies \citep{dhooge2006matcont}. Detailed bifurcation analysis of the single Jansen-Rit neural mass model, can be found in \citep{touboul2011neural}. To determine whether firing rate affects Jansen-Rit dynamics, Grimbert and Faugeras \citep{grimbert2006bifurcation} brought all the other parameters to the physiological values provided in Table 1. Moreover, 2-codimension bifurcation in a single Jansen-Rit model were investigated as a function of two critical parameters and a very rich bifurcation diagram was presented \citep{touboul2011neural}.
A codimension 2 bifurcation diagram is plotted in figure \ref{bif_1}. A positive or negative value of these parameters ($ \alpha $ and $ f_{0} $) is mathematically permissible. Although negative inputs are not fully physiologically relevant, they provide an insight into the origin of dynamical features observed at more physiologically relevant values. The light blue color region indicates the steady-state activity, and the solution will converge to this fixed point regardless of the initial conditions.
The black star in the diagram represents Generalized Hopf (GH) bifurcation ($ (\alpha, f_{0}) = (0.2275, 40.2397) $) resulting from the Bogdanov-Takens (BT) bifurcation. The green (orange) color curve shows the subcritical (supercritical) Hopf bifurcation that is related to the negative (positive) fixed-point.
The BT bifurcation point is on the lower branch of the saddle-node bifurcations curve (red color). This BT bifurcation, gives rise to subcritical Hopf bifurcation (green curve) or a curve of saddle homoclinic bifurcations (purple curve).
The red curve illustrates a saddle-node bifurcation generated by a Cusp bifurcation (CP). The region between the purple and red color border (pink zone) displays spiking behavior with a large amplitude (figure \ref{bif_1}(\textbf{c})) and low frequency ($ \sim $ 4 Hz). In the purple border, fixed-point state and oscillatory state are separated by Hopf cycles. In ($ \alpha $, $ f_{0} $) = (0, 120) large amplitude signals are appear. Results show that when ($ \alpha $, $ f_{0} $) = (0, 130), spiking behavior will disappear and the system emerges oscillatory activities. There is an oscillatory activity in the green region that is associated with alpha rhythms (figure \ref{bif_1}(\textbf{d})).

\subsection{Analysis of network behavior based on nonidentical input/output from each node:}
\begin{figure}[!ht]
\centering
\includegraphics[width=10cm, height=4.5cm]{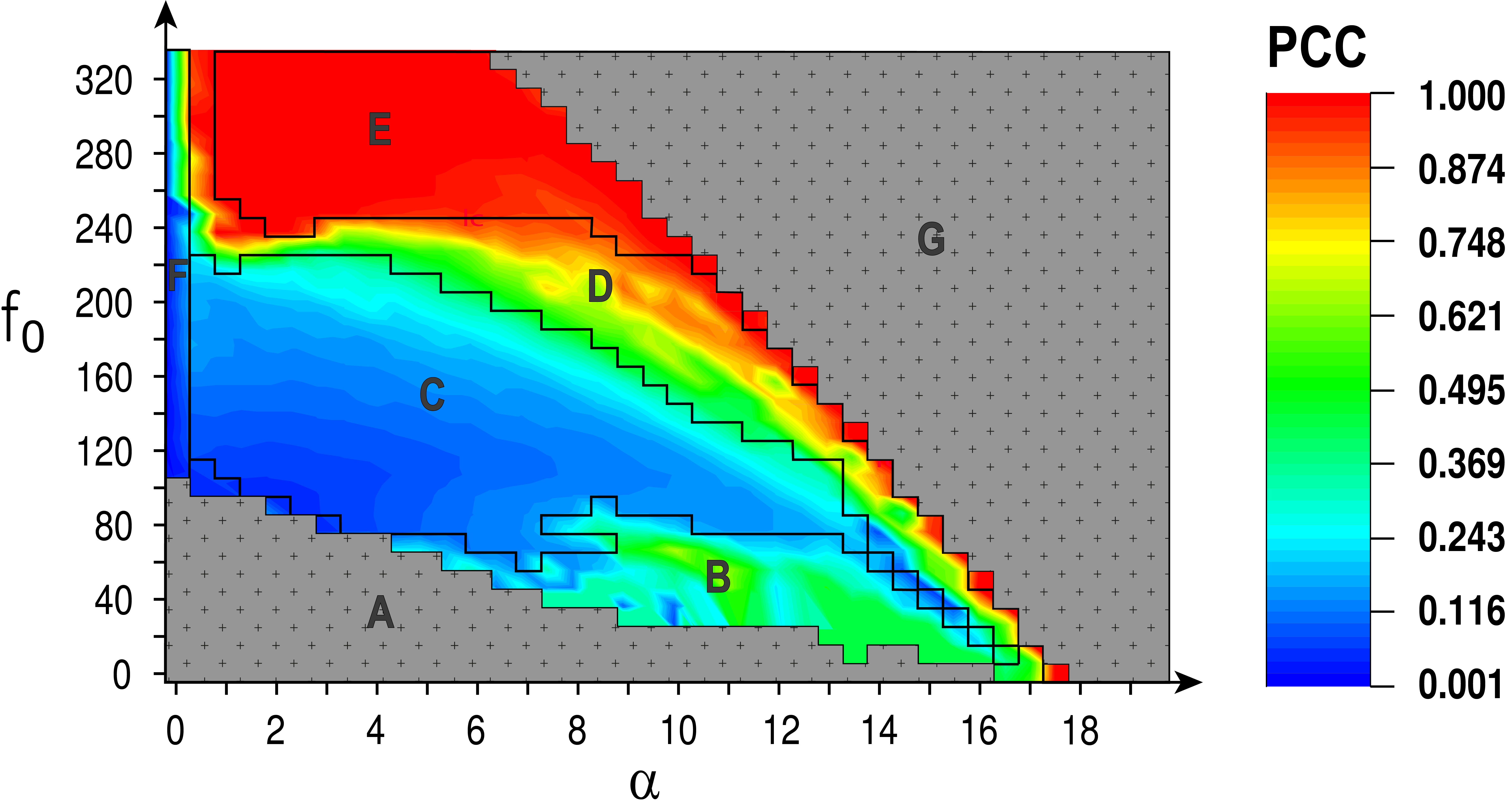}
\caption{The mean Pearson cross coefficient as a function of coupling strengths ($ \alpha $) and external inputs ($ f_{0} $). Each color and label exhibits different dynamical properties. This figure is divided into different segments based on the activity type and synchronization value.}
\label{corr}
\end{figure}

Although the simplicity of the mean-field approach is their strong point, using this assumption poses that all nodes are fully synchronized, and consequently the system is being at a full synchronization state, which is not the case in reality. So, we relax this assumption. The dimensionality and complexity of the system make it impossible to plot bifurcation diagrams with the Matcont package in this case. We construct a grid 
$ (\alpha, f_{0}) = \left\lbrace 0, 0.5, 1, 1.5, ..., 19.5\right\rbrace  \times \left\lbrace 0, 10, 20, 30, ...330\right\rbrace . $
Making use of the measure described in Section \ref{method}, we study the network synchronization. The scheme of the mean synchronization matrix as a function of coupling coefficients and external inputs is found in figure \ref{corr}. This figure is divided into different segments based on the activity type. All of these areas are thoroughly investigated as follows. 

\subsubsection*{G area:}
As a result of either the extremely high coupling strength or the external input, the system is at rest. Due to excessive input, it is reasonable for the system to reach a stable equilibrium. The numerical amount of fixed points is the positive number.

\subsubsection*{A area:}
\begin{figure}[!ht]
\centering
\includegraphics[width=10cm, height=4.2cm]{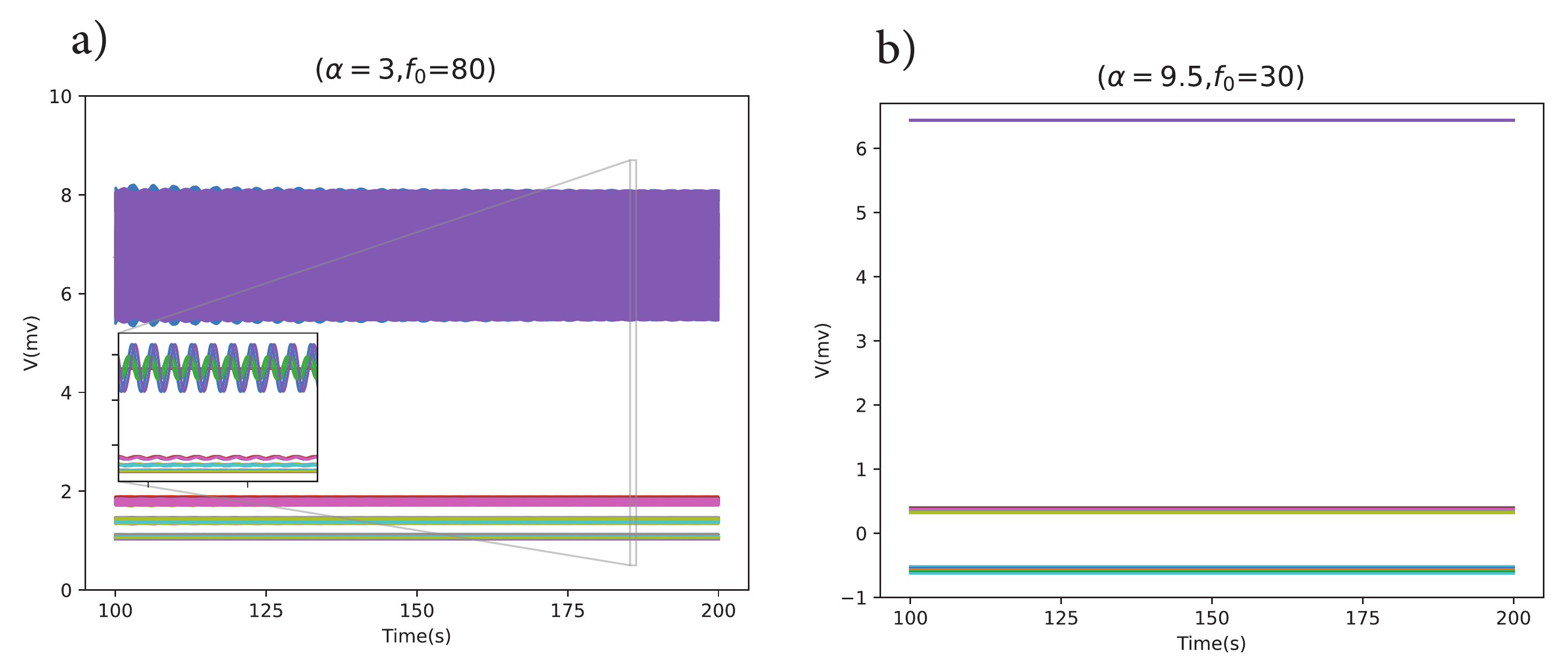}
\caption{Two behaviors in the border between 'A' area and 'B' and 'C' zones. Due to different initial conditions and different inputs during the simulation, some nodes are being at rest, whereas others emit oscillatory activity or take different fixed points (panel \textbf{(a)}). In some cases, all nodes are in a fixed-point state (panel \textbf{(b)}).}
\label{E_border}
\end{figure}

Contrary to the 'G' zone, the weak intensity of input cannot create fluctuations in the 'A' zone, which means that the system has entered a state of rest. Although the steady-state for small values of external input ($ < $50) is negative, which has no biological significance, the neighbors each of these has a wide variation of behaviors. Some interesting behaviors along the border between this area and the 'B' and 'C' areas are seen. It is observed that the system starts to fluctuate, and after a short time goes to rest.
It has indeed been unable to remain oscillating forever and has finally entered a fixed-point. Throughout this zone, multi-stability can exist between nodes. It means that due to different initial conditions and different inputs during the simulation, some nodes are being at rest, whereas others emit oscillatory activity or take different fixed points. Notably, the mean system activity finally reaches a stable equilibrium. As shown in figure \ref{E_border}, two cases have been observed in this area.

\subsubsection*{B area:}
\begin{figure}[!ht]
\centering
\includegraphics[width=10cm, height=10cm]{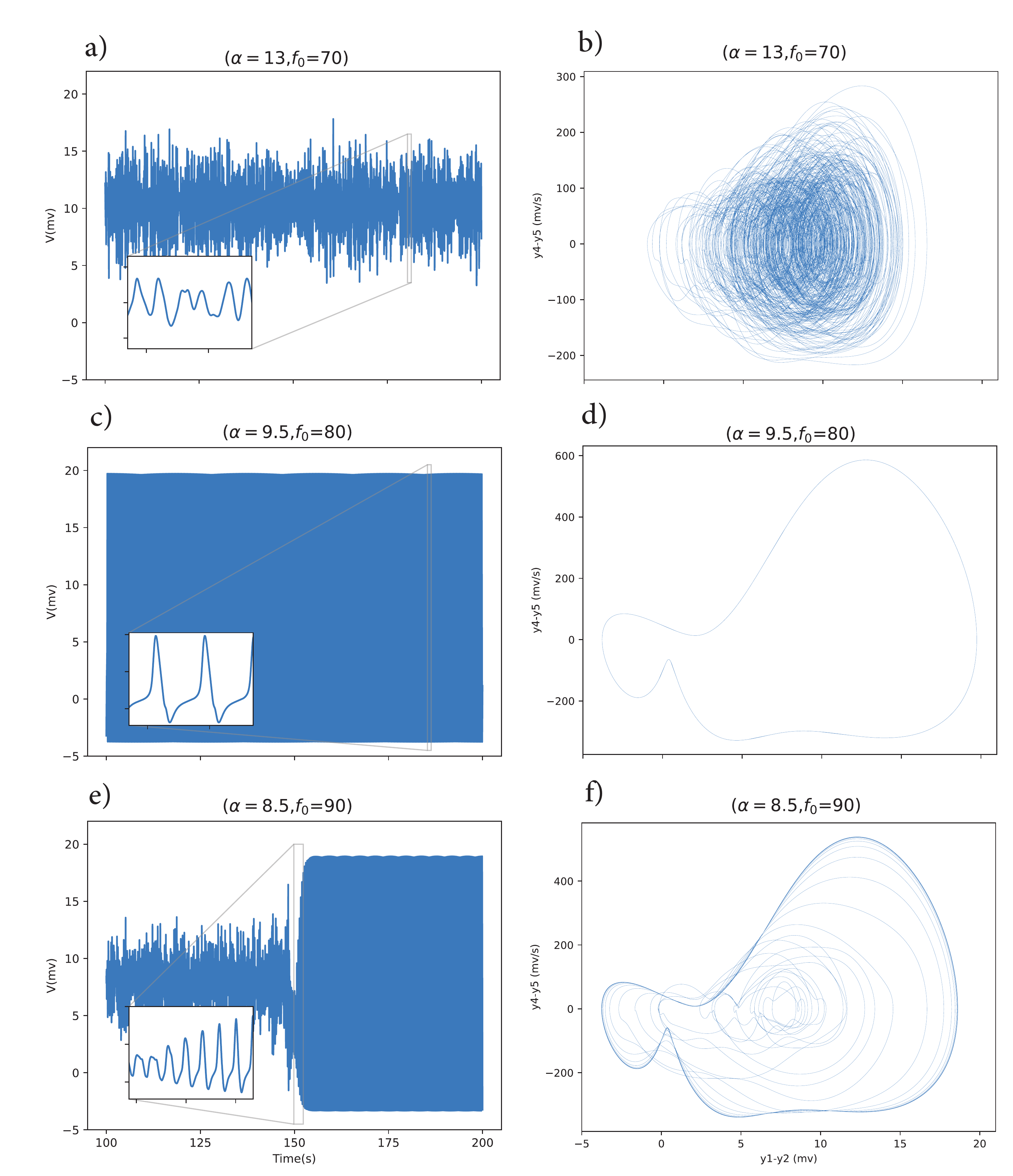}
\caption{Three behaviors in the 'B' area. Chaotic (\textbf{(a)}), periodic (\textbf{(c)}), and mixtures of these two oscillation states (\textbf{(e)}) are shown. The phase portraits of (\textbf{a}), (\textbf{c}), and (\textbf{e}) are viewed in (\textbf{b}), (\textbf{d}), and (\textbf{f}) respectively.}
\label{H_area}
\end{figure}

\begin{figure}[!ht]
\centering
\includegraphics[width=10cm, height=3.5cm]{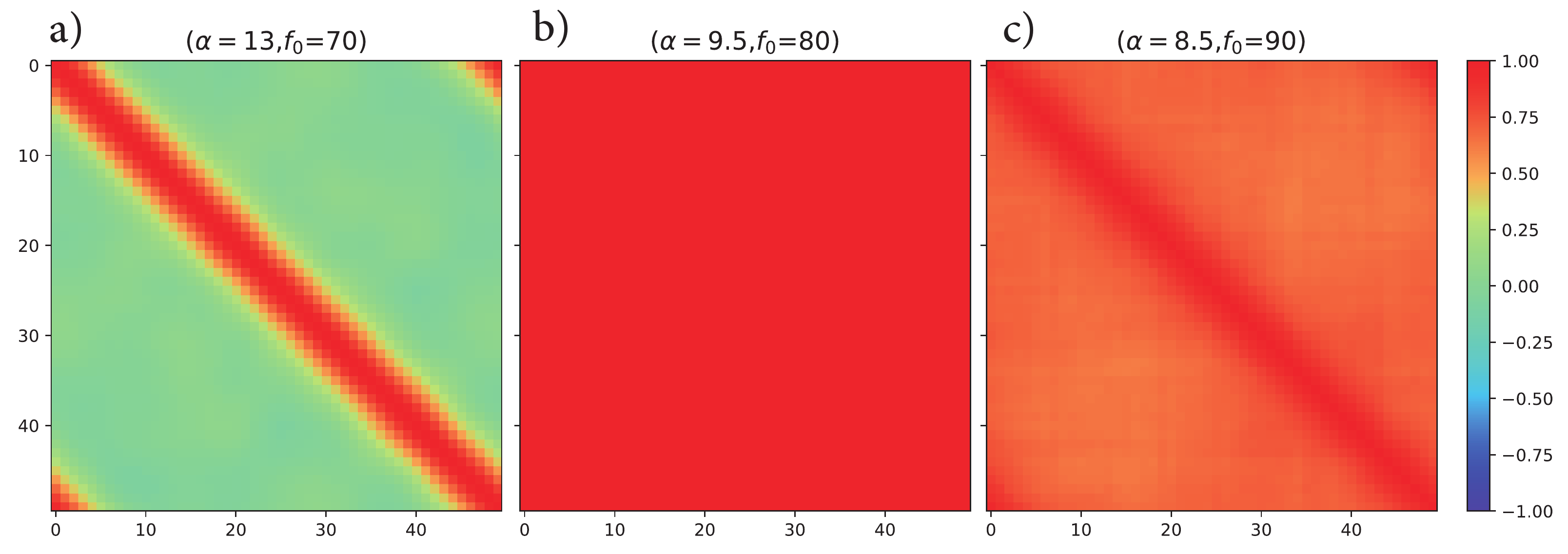}
\caption{The synchrony matrices of states from figure \ref{H_area}. (\textbf{a}) a zero correlation (in chaotic activity), (\textbf{b}) a complete correlation (in ordered periodic signals), and (\textbf{c}) a correlation approximately equal to 0.5 (in mixed of ordered and disordered activity) are exist.}
\label{corr_H}
\end{figure}

Individual oscillators leave their fixed points and start to fluctuate. Observe that when the single cell of Jansen-Rit dynamics receives input of over 315, it goes to rest \citep{grimbert2006bifurcation}. As the inserted inputs to each unit in this area are more than this value, we expected the system to be at rest, but spiking mode is observed.\\
Different types of activity, including periodic and nonperiodic spikes and chaotic activity with the low-frequency band, are visible in the 'H' area. A generalized spike wave discharge (GSWD) occurs in this region. In the corticothalamic networks, GSWDs typically occur after paroxysmal \citep{martinet2017human}, but their exact mechanism is not clear yet. The emergence of high amplitude signals confirms a disorder such as epileptic seizure in a brain network \citep{maturana2020critical}. In figure \ref{H_area}, the left column is related to the chaotic, periodic, and mixed of these two oscillation states, respectively, and their phase portraits are viewed in the right column. Results show that the mean and variance of the cross-correlation matrix are varied widely. Figure \ref{corr_H} depicts the synchrony matrices of states from figure \ref{H_area}. A zero correlation (in chaotic activity), a complete correlation (in ordered periodic signals), and a correlation approximately equal to 0.5 (in mixed of ordered and disordered activity) exist. Actually, a range of different patterns of the correlation matrix is found. Noteworthy, a saddle-node bifurcation has occurred between the 'A' and the 'B' areas at the border between fixed-point state and spiking activity.

\subsubsection*{C area:}
\begin{figure}[!ht]
\centering
\includegraphics[width=10cm, height=10cm]{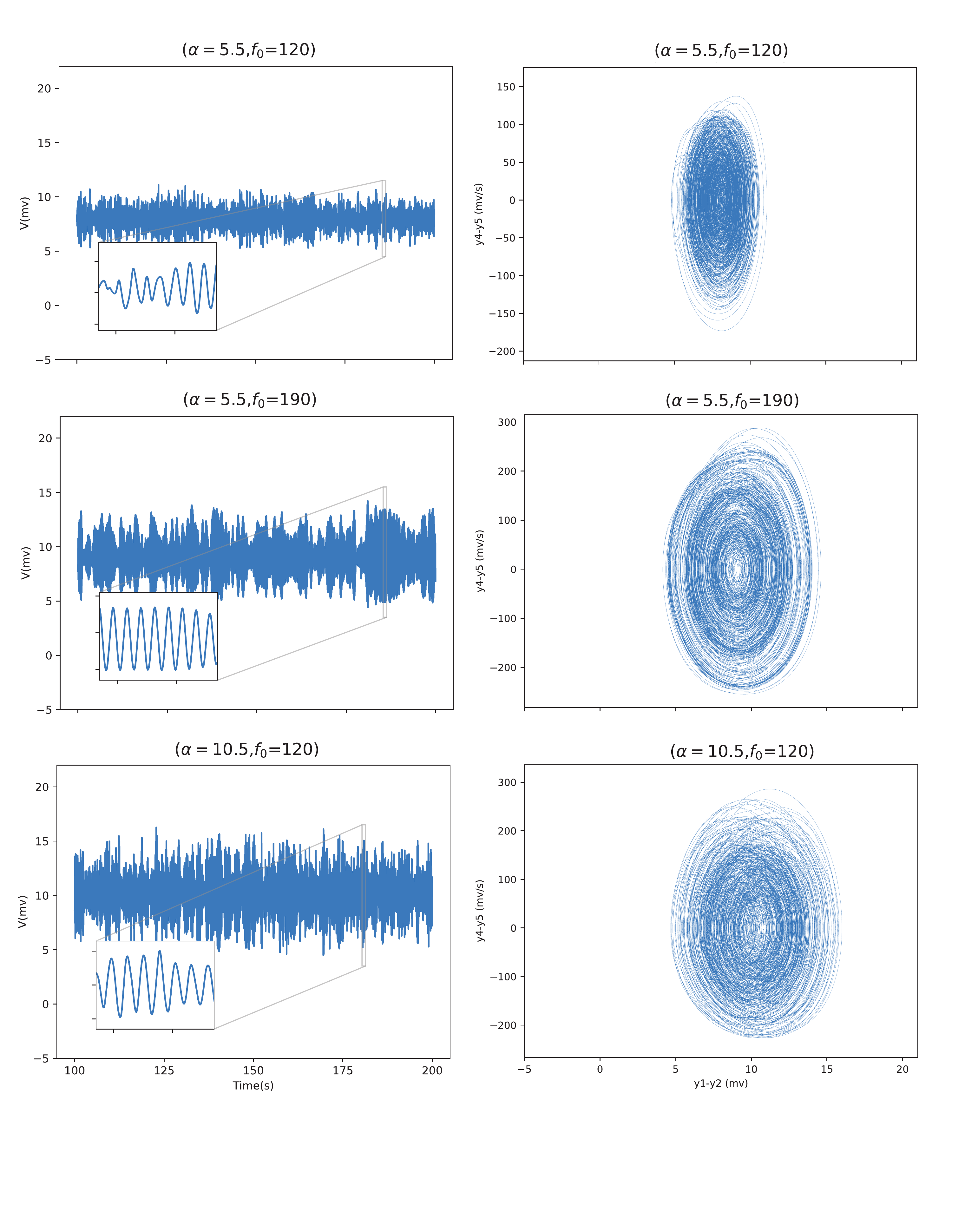}
\caption{Three arbitrary behaviors (left columns) and their phase portrait (right columns) in the 'C' area. Increased coupling coefficients or external input results in larger signal amplitudes and expanded phase-space volumes.}
\label{G_area}
\end{figure}

\begin{figure}[!ht]
\centering
\includegraphics[width=10cm, height=3.5cm]{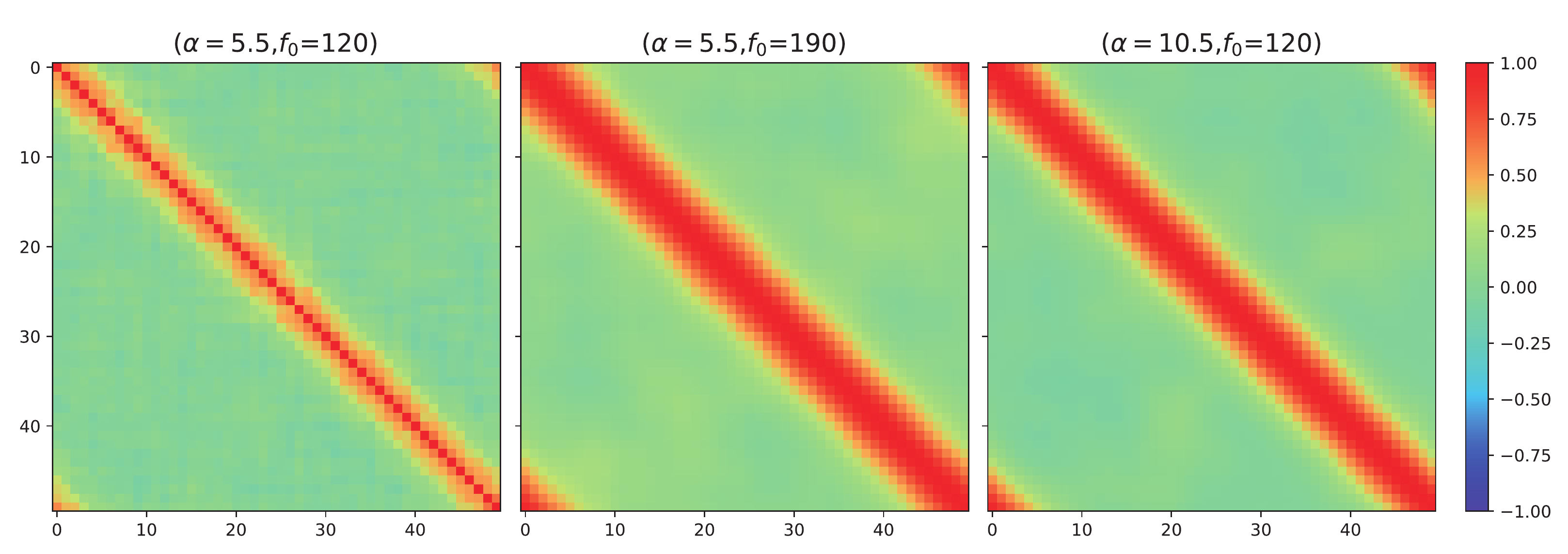}
\caption{The synchrony matrices of states from figure \ref{G_area}. The correlation value in this area is close to zero, which means there is no linear relationship between the nodes.}
\label{G_corr}
\end{figure}

All nodes exhibit oscillatory activity in region 'C'. The fluctuations have no fixed amplitude and have changed based upon the coupling coefficient or external input. Increasing or decreasing these factors have affected the fluctuation amplitude. Three arbitrary states in this regime and their phase portrait are shown in the left and right columns of figure \ref{G_area}, respectively. Figure \ref{G_corr} shows the synchrony matrix for each of the three cases in figure \ref{G_area}. Increased coupling coefficients or external input results in larger signal amplitudes and expanded phase-space volumes. The correlation value in this area is close to zero, which means there is no linear relationship between the nodes. Chaos is the name we give to this area. The boundary between 'A' and 'C' refers to the appearance of a periodic orbit through a change in the stability properties of a fixed point known as the Hopf bifurcation. Moreover, taking synchronization as an order parameter, this boundary can separate the ordered phase (A) from the disordered phase (C), and so, a phase transition is expected.

\subsubsection*{D area:}
\begin{figure}[!ht]
\centering
\includegraphics[width=8cm, height=8cm]{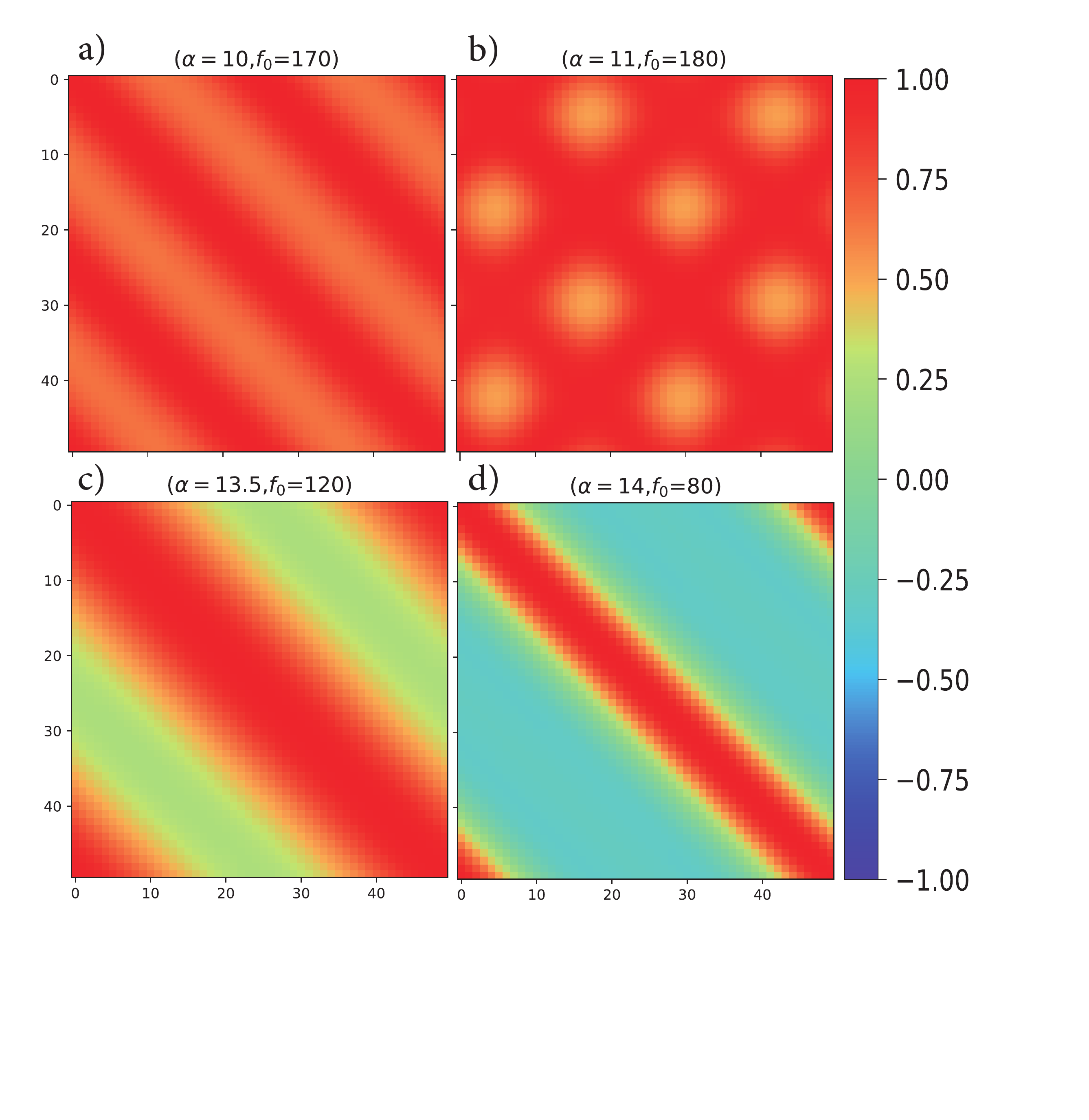}
\caption{Four different patterns in correlation matrices in the 'D' zone. In the border of the 'E' area (panel \textbf{(b)}, \textbf{(c)}), the correlation matrix has a high mean. There is a low mean in the correlation matrix in the 'C' area (panel \textbf{(d)}).}
\label{2_corr}
\end{figure}

\begin{figure}[!ht]
\centering
\includegraphics[width=8cm, height=11cm]{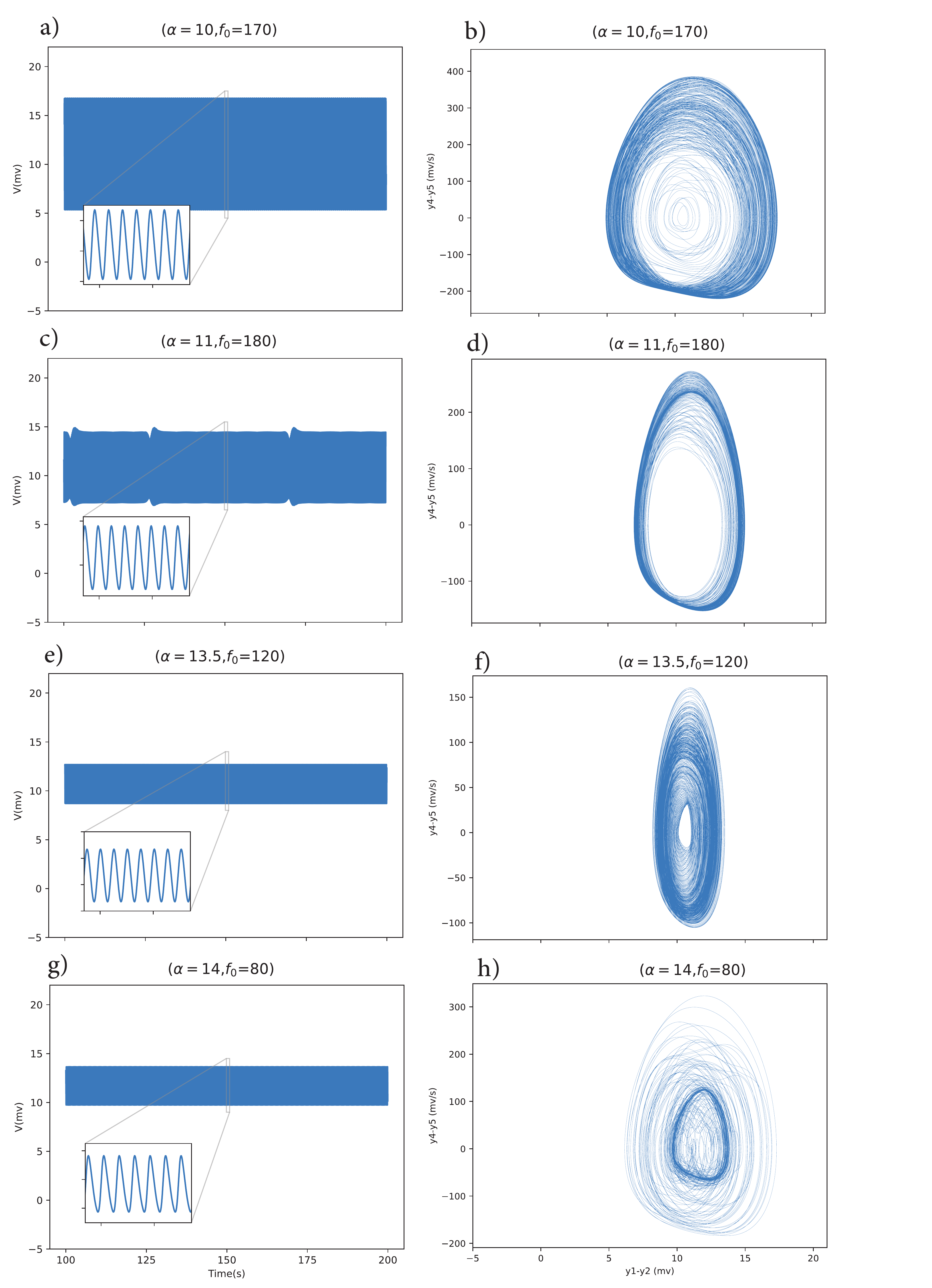}
\caption{The time series (left columns) and their phase spaces (right columns) of states from figure \ref{2_corr}. The modulated signals are visible and there is no spiking activity in this region (left columns). Panel \textbf{(d)} and \textbf{(f)} are on the border of the 'E' area, and they show a phase space close to that of a regular limit cycle. On the edge of the 'C' area, irregularity trends are observed in panel \textbf{(h)}. Panel \textbf{(b)}  is composed of regular and irregular activities because possessing no common border with any area.}
\label{2_area}
\end{figure}

There is no spiking activity in this region, and the modulated signals are visible. Different exciting patterns in correlation matrices have emerged. Indeed, the 'D' zone is located between full synchronous ('E') and asynchronous ('C') areas. A summary of the most repeatable is shown in figure \ref{2_corr}. On the border of the 'E' ('C') area, the correlation matrix has a high (low) mean. Figure \ref{2_area} shows the mean signals and their phase spaces for each matrix in figure \ref{2_corr}. $ (\alpha, f_{0}) = $ (13.5, 120), $ (\alpha, f_{0}) = $ (11, 180) are in the border of 'E' area and are showing a phase space close to that of a regular limit cycle. Also, $ (\alpha, f_{0}) = $ (13.5, 120) displays periodic and regularity patterns. On the edge of 'C' area, irregularity trends are observed in $ (\alpha, f_{0}) = $ (14, 80). In $ (\alpha, f_{0}) = $ (10, 170) that possesses no common border with any area, the phase space is composed of both regular and irregular activity.

\subsubsection*{E area:}
\begin{figure}[!ht]
\centering
\includegraphics[width=12cm, height=3.7cm]{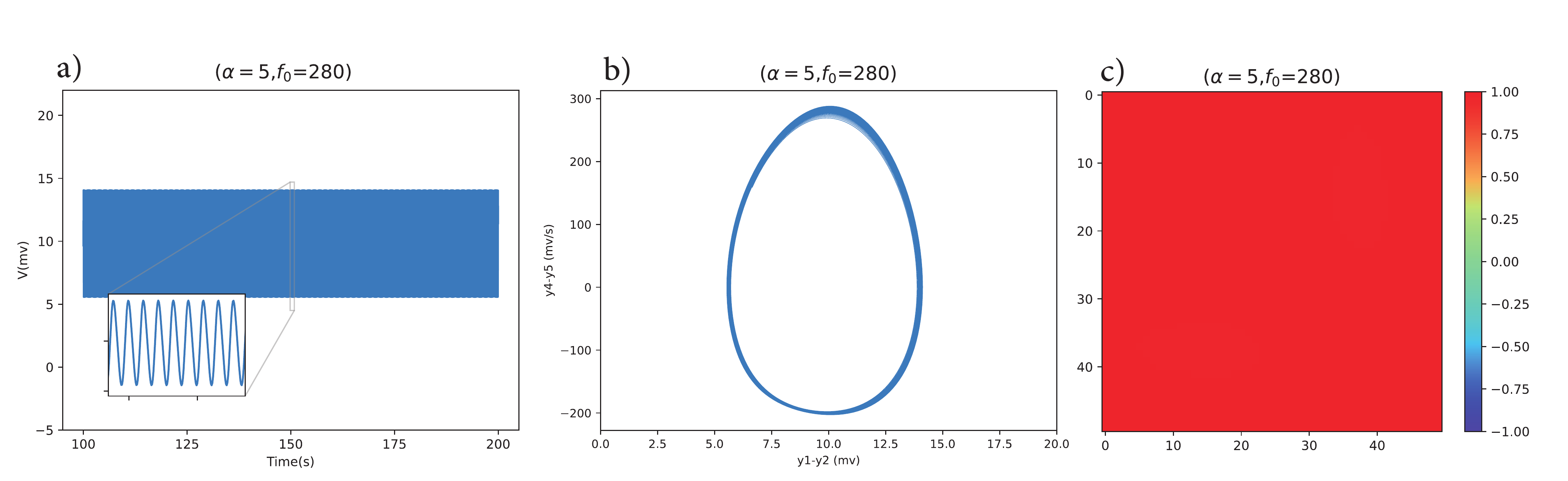}
\caption{One arbitrarily chosen sample of the mean signal (panel \textbf{(a)}), its phase space (panel \textbf{(b)}), and its correlation matrix (panel \textbf{(c)}) in the 'E' region. The nodes are approximately synchronized, and complete system synchronization is observed.}
\label{F_all}
\end{figure}

Here is where the complete system synchronization begins and the nodes are approximately synchronized. Indeed, the system tends to synchronize strongly with each coupling coefficient. A single Jansen-Rit mass possesses a stable equilibrium where the received input exceeds 315 \citep{grimbert2006bifurcation}.
Surprisingly, almost all of the input has received by the system in this zone is more than 315, but the system still shows oscillatory behavior. Actually, interactions between nodes make this possible. The one arbitrarily chosen sample of the mean signal, its phase space, and its correlation matrix in this area are displayed in figure \ref{F_all}. 

\subsubsection*{F (Zero coupling coefficient) area:}
\begin{figure}[!ht]
\centering
\includegraphics[width=10cm, height=3.8cm]{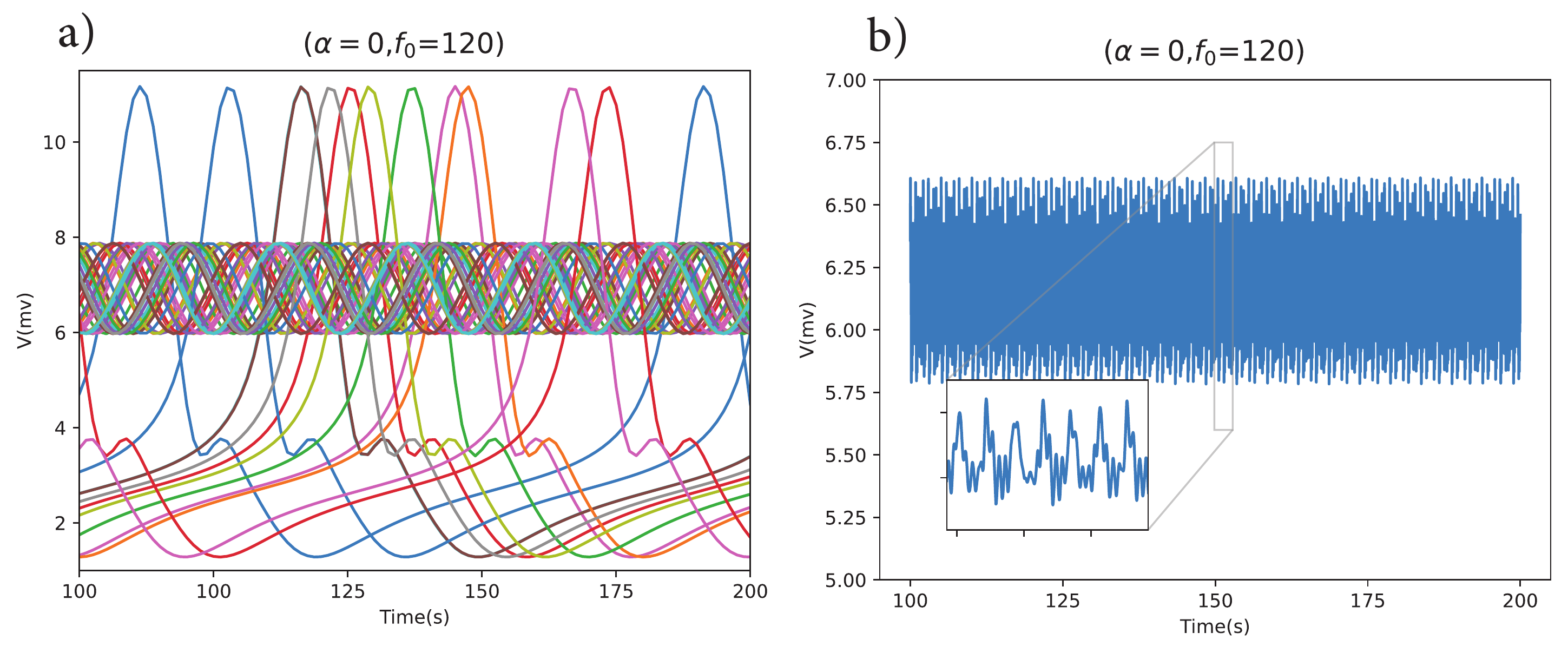}
\caption{(\textbf{a}) The activity of each node for $ (\alpha, f_{0}) $ = (0, 120).  Two Spiking and Oscillating activities are visible in each node. Panel (\textbf{b}) shows the mean of these activities.}
\label{res_120}
\end{figure}

In uncoupled oscillators ($ \alpha = $ 0 in the system) with external input less than 90, each node is at a fixed point. So, the mean network behavior is non-active. Note that the value of system fixed-point is negative (positive) for external input less (more) than 50.\\
In $ f_{0} = $100, 110, some nodes start to fluctuate. Each node has two states in these values: a fixed-point and a stable limit cycle producing the alpha rhythm.
The value of $ f_{0} = $ 120 is the minimum of external input at which all nodes leave the fixed-point and show fluctuations. Moreover, two types of activity are visible in each node: Spiking and Oscillating (figure \ref{res_120}\textbf{(a)}). Therefore, the appearance of the mean activity has a strange trend (figure \ref{res_120}\textbf{(b)}).
This behavior is continued up to $ f_{0} = $ 140. The spiking type disappears at the higher value of $ f_{0} = $ 140. When $ f_{0} $ varies between 140 and 310, only oscillatory activity occurs at individual nodes. At $ f_{0} = $ 320, all nodes leave the limit cycle and rest in a fixed point. \\
As a summary, two Hopf bifurcations occur at $ f_{0} \simeq $ 90, 320. At $ f_{0} \simeq $ 120, a saddle-node on an invariant curve (SNIC) bifurcation happens. Moreover, at $ f_{0} \simeq $ 140, a Limit Point of Cycle (LPC) bifurcation is observed.
It is imperative to note that the zero coupling coefficient region is the same as a single Jansen-Rit mass that has been investigated in detail in \citep{grimbert2006bifurcation}.

\subsubsection*{Some special cases:}
Each segment of the synchronization matrix map has been investigated based on its properties so far. We have examined in-dept several specific characteristics such as the time series of each unit and the correlation matrix between them, the mean signals, and finally, their phase portraits of the network.
It is important to note that some interesting patterns are formed at a boundary between areas. Several of these cases are examined in detail as follows:\\
\begin{itemize}
\item $ (\alpha, f_{0}) $ = (17.5, 0): At this point, the nodes are synchronized, and the network shows either a resting state at the negative fixed-point or oscillatory activity depending on initial conditions. This coexistence of two distinct attractors at the same parameter value is called bistability.  According to theoretical and empirical studies, bistability is associated with first-order phase transitions \citep{cowan2016wilson, cocchi2017criticality}. Mathematically,  this transition is equivalent to subcritical bifurcation \citep{pomeau1986front}. Indeed, the mean signal of the model erratically switches between a fixed point and a limit cycle attractor. The nodes are either in a quiescent phase or an oscillating state simultaneously (Figure \ref{Fig16}). The phenomenon of bistability can be explained very well by a type of bifurcation known as the cusp \citep{harlim2007cusp}.

\begin{figure}[!ht]
\centering
\includegraphics[width=10cm, height=3.8cm]{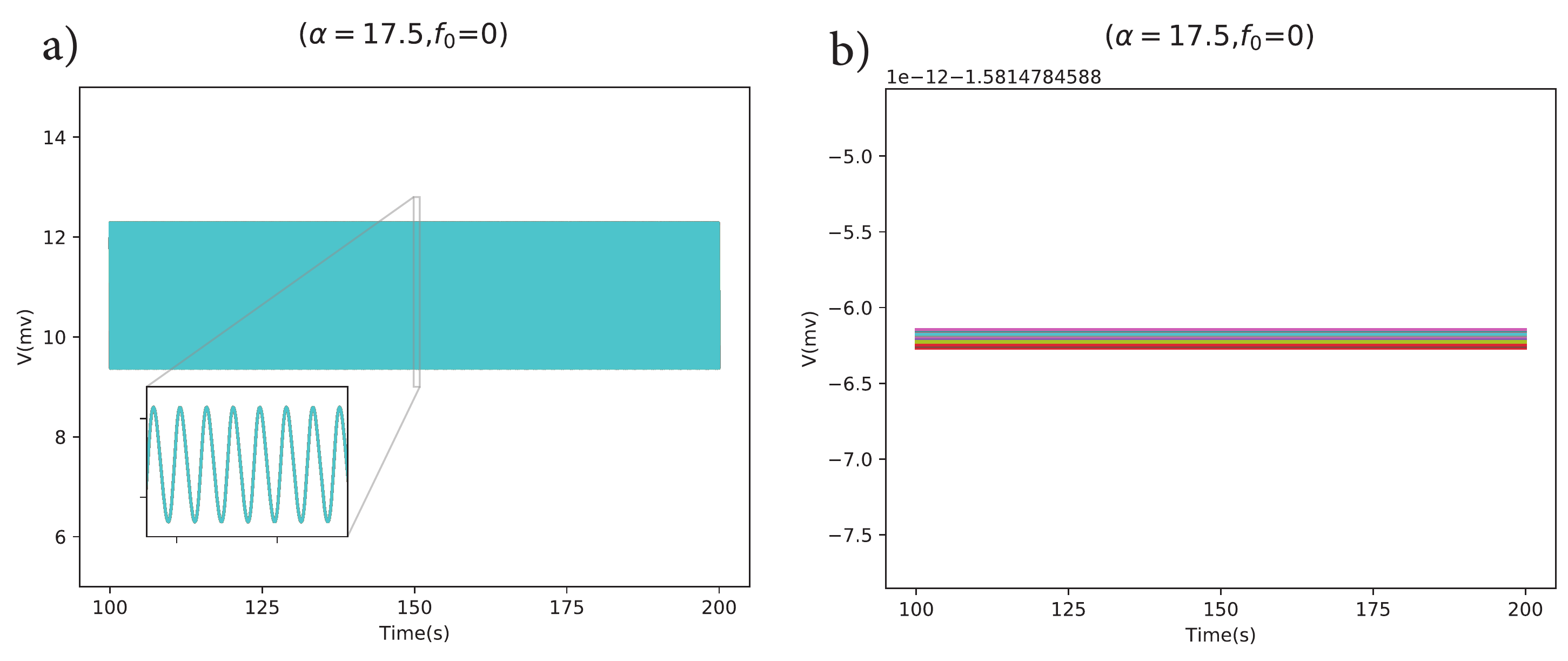}
\caption{Two possible behaviors for $ (\alpha, f_{0}) $ = (17.5, 0).  At this point, the nodes are synchronized, and the network shows either a resting state at the negative fixed-point (panel \textbf{(b)}) or oscillatory activity (panel \textbf{(a)}) that depends on initial conditions. Indeed, the nodes are either in a quiescent phase or an oscillating state simultaneously.}
\label{Fig16}
\end{figure}

\item $ (\alpha, f_{0}) $ = (2, 90), (1, 100): The three types of behavior presented here by each node are fixed-point state, spiking activity, or oscillatory activity. The mean of the network either exhibits a fixed point attractor or spiking activity. Figure \ref{special_case} shows two arbitrary runs of each state. Moreover, there are numerous patterns in the Pearson correlation matrix (Figure \ref{special_case_corr}). The red (blue) color indicates full synchronization (anti-synchronization). Local and global synchronization is also shown in these figures. These types of synchronizes are denoted by the red and blue masses, respectively. In local synchronization, some ensembles of neurons behave the same,  i.e., synchronized clusters, but others act in completely different ways. In fact, Pearson cross-correlation matrices containing only red color signify a global synchrony state. The blue and red masses indicate in-phase and antiphase local synchrony states, respectively. Additionally, The presence of green masses is related to local asynchrony, which means there is no synchronization. This different pattern could reflect different brain network synchronization patterns during different types of tasks \citep{nazemi2019influence}.\\
Due to the high variance of the synchronization value, these points are suitable candidates to identify the second phase transitions. In order to claim that the second phase transition occurs at any of these points, we should compute the coefficient of variation (CV) against the control parameter \citep{di2018landau}. Indeed, CV is a statistical measure of the dispersion of data points in a data series around the mean. A peak in the coefficient of variation during the continuous phase transition (if it exists) would be the first marker of the second phase transition \citep{kazemi2022phase}.
According to figure \ref{CV}, $ (\alpha, f_{0}) $ = (2, 90), (1, 100) can be shown the second phase transition. Notably, the CV value for alpha greater than 3 at f=90, 100 is approximately zero and therefore is discarded in figure \ref{CV}.

\begin{figure}[!ht] 
\centering
\includegraphics[width=9cm, height=7.5cm]{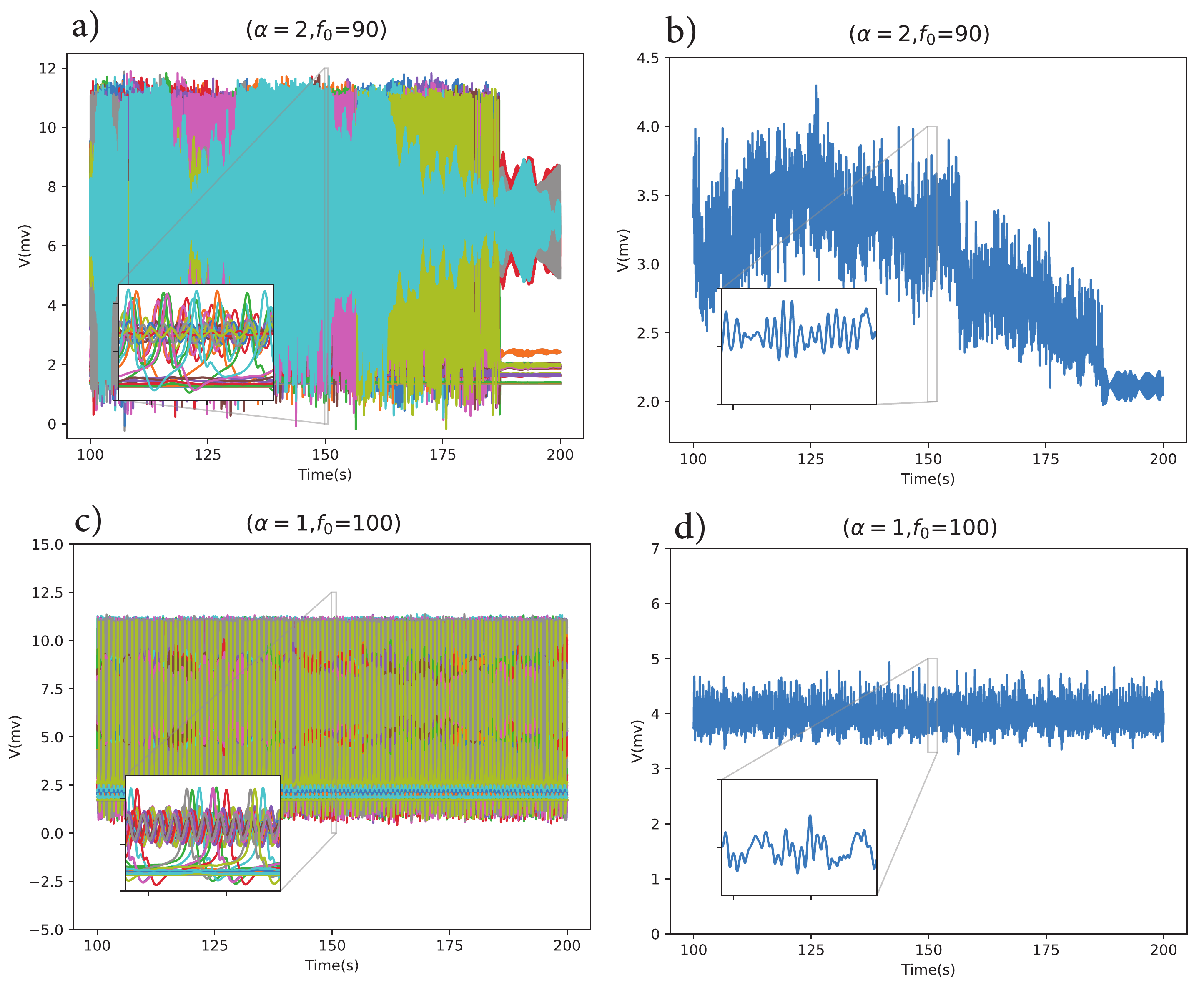}
\caption{Two arbitrary runs for $ (\alpha, f_{0}) $ = (2, 90), (1, 100). The three types of behavior by each node are fixed-point state, spiking activity, or oscillatory activity (panel \textbf{(a)} and \textbf{(c)}). Panel \textbf{(b)} and \textbf{(d)} show the mean activity of \textbf{(a)} and \textbf{(c)}, respectively.}
\label{special_case}
\end{figure}

\begin{figure}[!ht]
\centering
\includegraphics[width=7cm, height=5.7cm]{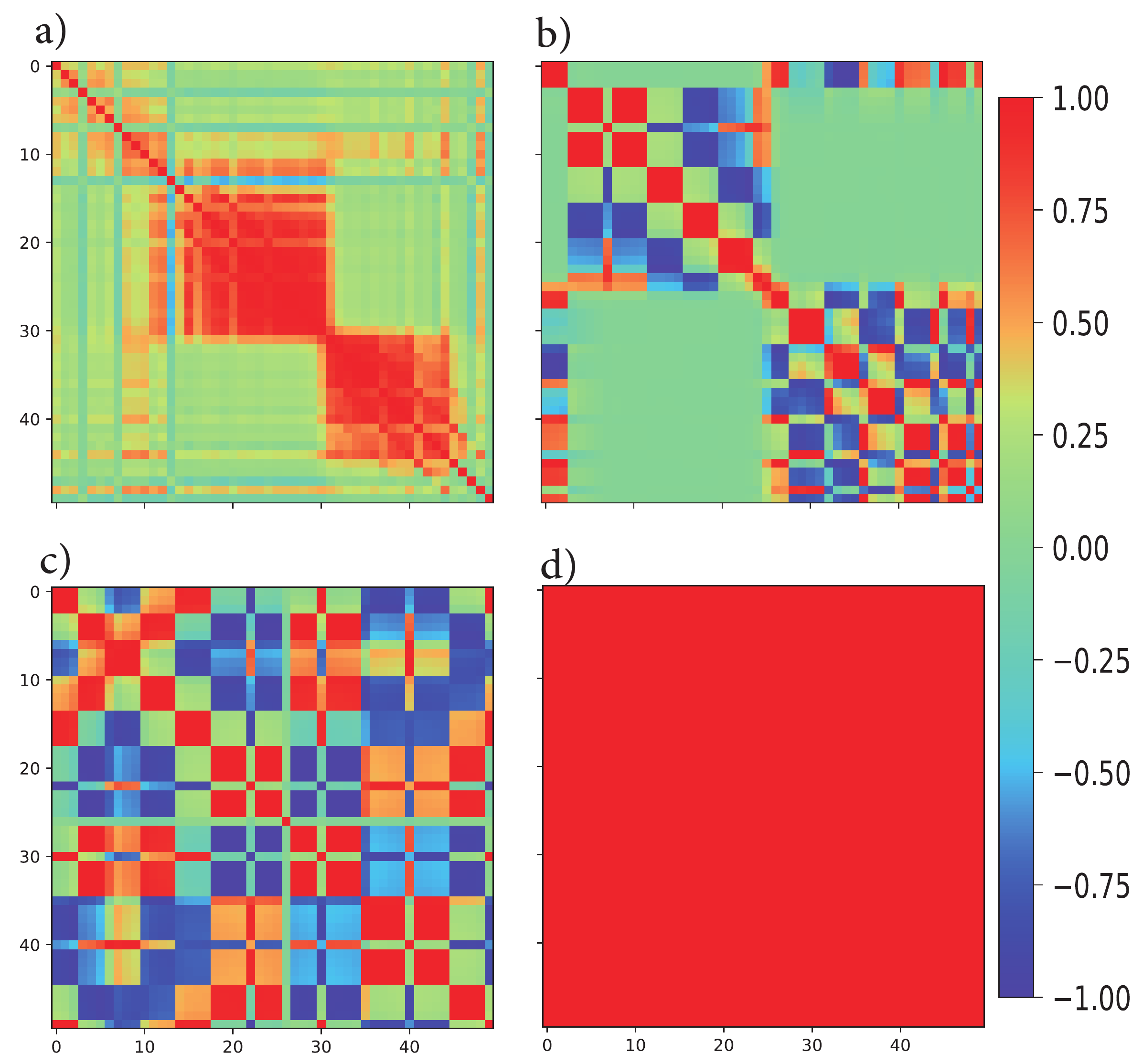}
\caption{Four patterns in the Pearson correlation matrix from figure \ref{special_case}. The red (blue) color indicates full synchronization (anti-synchronization). The green masses are related to local asynchrony, which means there is no synchronization (panel \textbf{(a)} and \textbf{(b)}). Some ensembles of neurons behave the same,  i.e., synchronized clusters that are visible in panel \textbf{(a)}. Panel \textbf{(c)} shows the asynchrony state. Panel \textbf{(d)} contains only red color signifies a global synchrony state.}
\label{special_case_corr}
\end{figure}

\begin{figure}[!ht] 
\centering
\includegraphics[width=8cm, height=6cm]{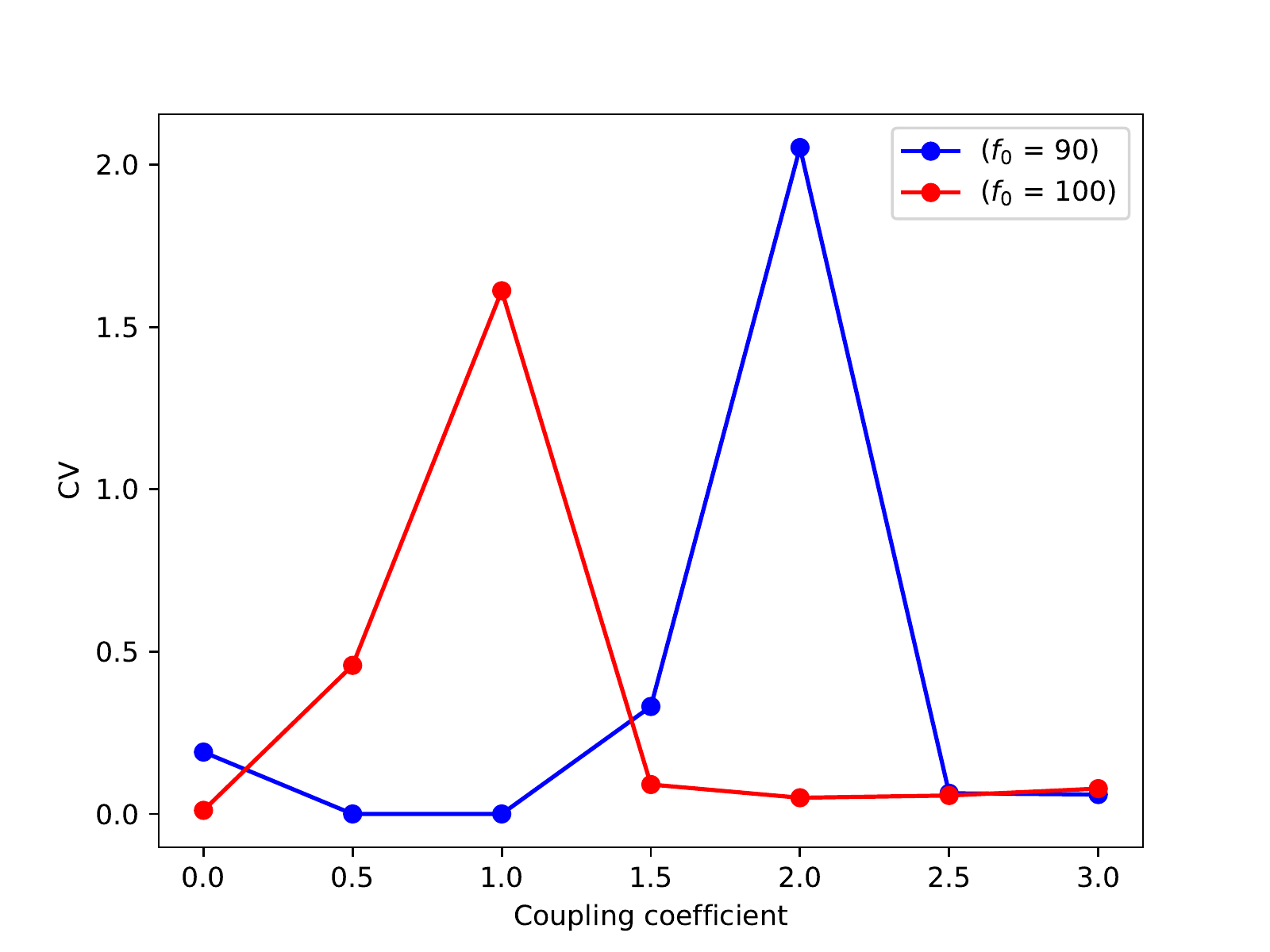}
\caption{The coefficient of variation (CV) against the coupling coefficient for $ f_{0} $ = 90, 100. These points of a maximum of these curves correspond to the value of the transition points. Notably, the CV value for alpha greater than 3 at f=90, 100 is approximately zero and therefore is discarded here.}
\label{CV}
\end{figure}

\item $ (\alpha, f_{0}) $ = (0.5, 110), (1.5, 100): Nodes here exhibit either spiking activity or oscillatory activity (Figure \ref{special_case_second}). The mean of the network shows the chaos caused by spiking and oscillation. The correlation matrix is similar to the 'G' area and has a zero value. 

\begin{figure}[!ht] 
\centering
\includegraphics[width=10cm, height=8.5cm]{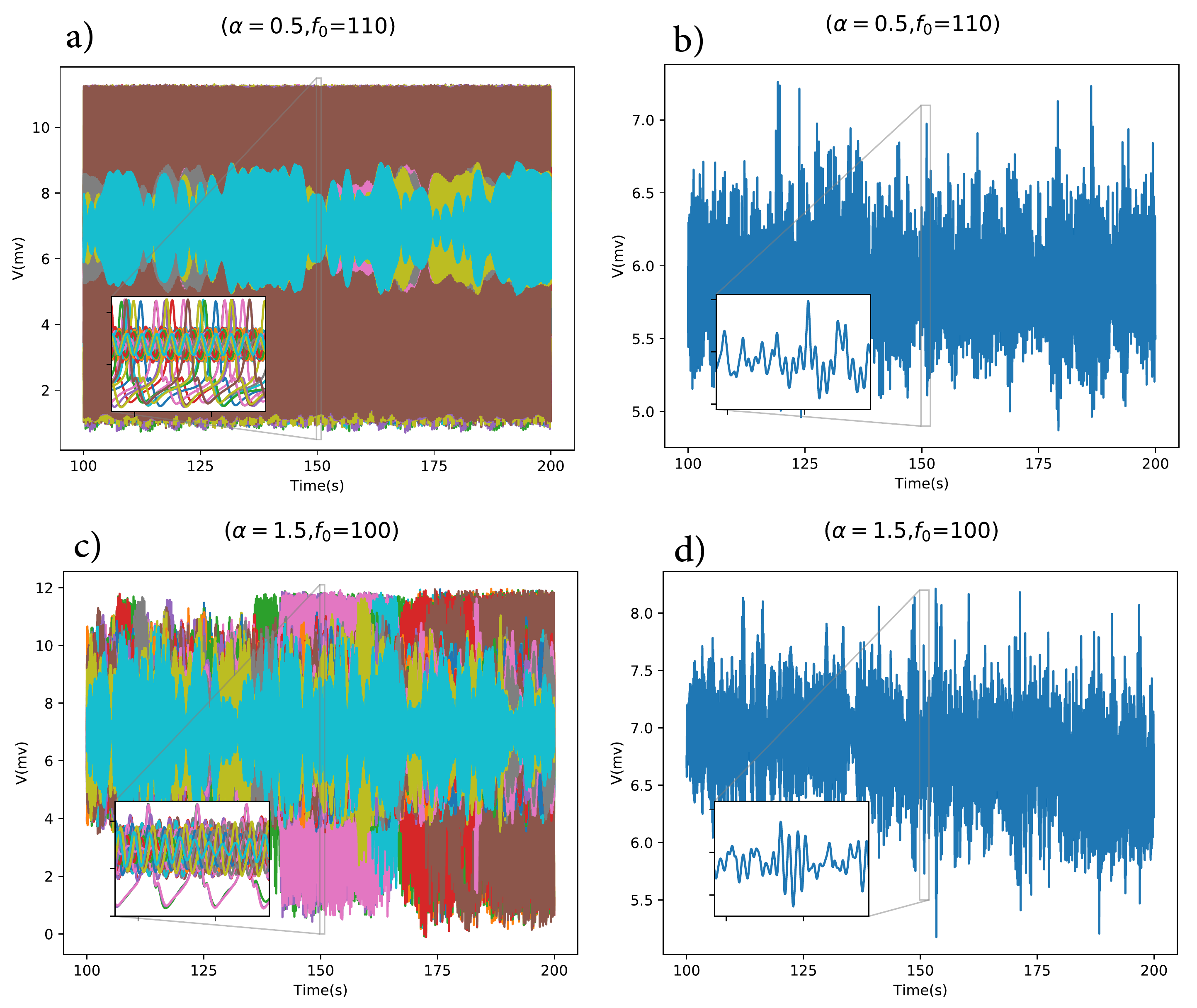}
\caption{Two arbitrary runs for $ (\alpha, f_{0}) $ = (0.5, 110), (1.5, 100). Nodes exhibit either spiking activity or oscillatory activity (panel \textbf{(a)} and \textbf{(c)}). Panel \textbf{(b)} and \textbf{(d)} show the mean activity of (a) and (c), respectively.}
\label{special_case_second}
\end{figure}

\end{itemize}
Interestingly, the first special case lies on the border of several different areas ('B', 'E' and 'G') and the first phase transition. So, these overlaps have been lead to different surprising results. Other special cases explained before are situated on the border of fixed-point and oscillatory activity that some of them are related to the second phase transition.

The correlation matrix between nodes, mean time series, and their phase spaces for all considering coupling coefficients and external inputs are available in Supplementary 1, Supplementary 2, and Supplementary 3, respectively.

\subsection{Frequency domain analysis}
\begin{figure}[!ht]
\centering
\includegraphics[width=13.2cm, height=5.2cm]{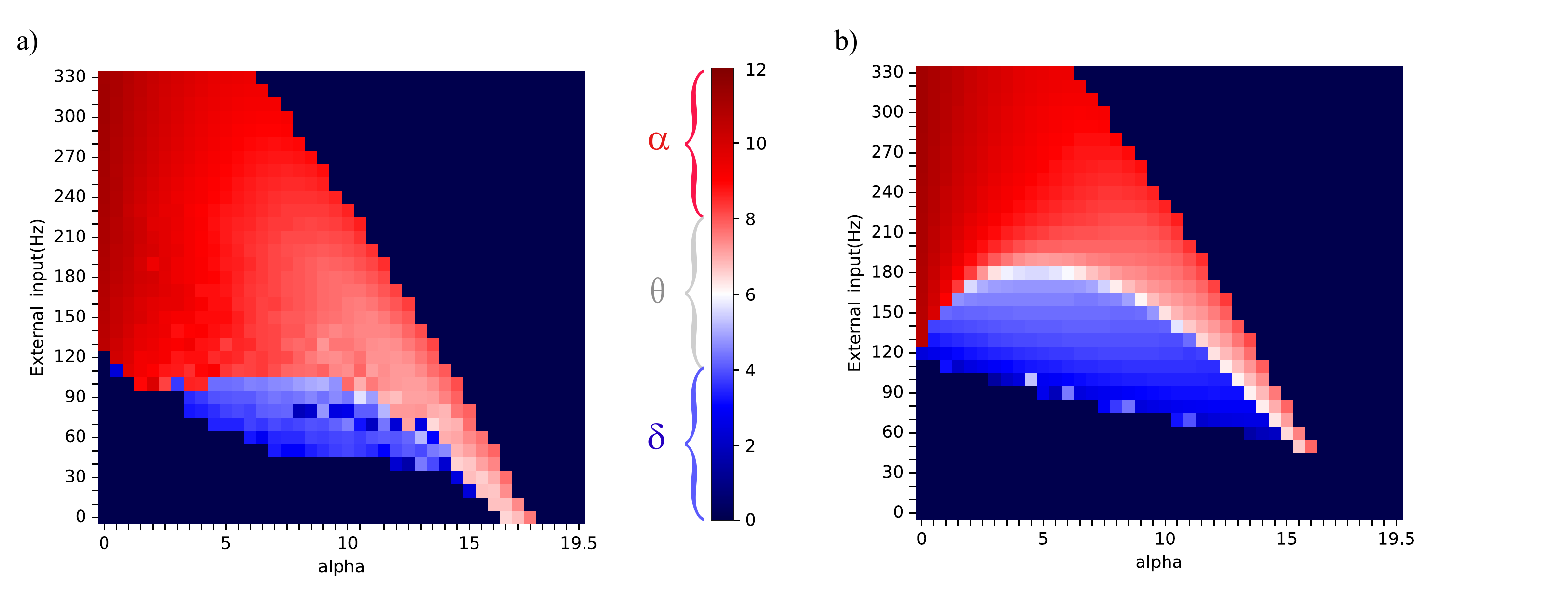}
\caption{The dominant frequency value as a function of the coupling term and external input. Panel \textbf{(a)} and \textbf{(b)} are related to nonidentical and identical inputs from each node, respectively. Different rhythms, including delta, theta, and alpha, can be observed in these two states. Moreover, an increase in one direction (coupling term or external input) increases the dominant frequency in \textbf{(a)} and \textbf{(b)}. Using Seaborn, a Python data visualization library based on Matplotlib, we visualized the data.}
\label{dominant rhythm}
\end{figure}

Using dominant frequency as a tool for signal analysis can provide valuable insight into brain disorders \citep{goel1996dominant, goossens2017eeg, newson2019eeg}. In order to determine the dominant frequency, the power spectrum of a signal is plotted. The dominant frequency is usually the one that carries the most energy, which can be observed on a magnitude spectrum as the peak frequency \citep{think2014digital, unpingco2016python}. Additionally, the fundamental frequency is the lowest frequency on a power spectrum with a peak \citep{telgarsky2013dominant}.
The dominant frequency value is plotted as a function of the coupling term and external input in figure \ref{dominant rhythm}. These subplots (a) and (b) are related to nonidentical and identical inputs from each node, respectively. (a) has a smaller steady-state region than (b). In other words, each node's nonidentical inputs contribute to the emergence of oscillatory behavior. Notably, an increase in one direction (coupling term or external input) increases the dominant frequency in (a) and (b) states. 

Different rhythms, including delta, theta, and alpha, can be observed in every two states, (a) and (b).

Delta oscillations correspond to the low-frequency range ($ < $ 4 Hz). Delta waves represent the slowest recorded brain waves with high amplitudes in humans and are associated with deep sleep stages. The presence of delta waves is prominent in brain injuries and learning disabilities \citep{alfimova2007changes}. They also affect concentration \citep{abhang2016introduction}. Moreover, the absence seizure is characterized by spike-wave discharges that occur during the delta-band frequency \citep{hindriks2013phase}. Due to figure \ref{dominant rhythm}, by altering the inputs in a network of Jansen-Rit neural mass models, delta waves can be generated that have not previously been observed in a single Jansen-Rit neural mass analysis \citep{grimbert2006bifurcation}. It is an emergent property of a complex system. The following statement states that no isolated node represents a delta rhythm, but this rhythm can emerge when these units interact with each other in a complex system.\\
Learning and memory may be affected by theta oscillations, a rhythm that appears at 4$ - $8 Hz \citep{seager2002oscillatory, eggermont2021brain}. An electrode recording in the hippocampal region can clearly reveal theta oscillations, which are high-amplitude neuronal oscillations \citep{buzsaki2002theta, munro2019theta}. Past studies have attributed theta rhythms to various clinical conditions, including epilepsy \citep{douw2010epilepsy}, however, their common occurrence has been recognized as a normal phenomenon \citep{mari2019normal}. Based on figure \ref{dominant rhythm}, nonidentical inputs (a) result in a smaller theta frequency region than identical inputs (b). Indeed, the alpha rhythm in (a) is propagated sooner than (b), which is more intriguing.\\
There are macroscopic oscillations of the brain in the range of 8$ - $12 Hz known as alpha waves. Analyses of electrophysiological methods, such as electroencephalography (EEG) and magnetoencephalography (MEG), detect the alpha waves as part of the brain wave activity \citep{fumagalli2012functional, ince2021inventor}. In the human brain, alpha-band oscillations are dominant and also play a significant role in information processing \citep{klimesch1997eeg, sigala2014role}. The first pixel in (b) that alpha rhythm can emerge, is in $ f_{0} $ = 50. Surprisingly, interacting nodes make it possible to observe the alpha rhythm even if there is no external input (a).

\section{Conclusion}
We studied the dynamical properties of a network composed of 50 coupled oscillators with Jansen-Rit masses under a regular Watts-Strogatz model. Each node receives both deterministic external and internal inputs based on what it receives from its neighbors.
We can use both the external input and coupling gain (internal input) as bifurcation parameters to obtain a more comprehensive understanding of network behavior.
Two fundamental forms of system output have been discussed: (i) using a mean-field approximation to generate identical input for all nodes and (ii) nonidentical input for all nodes.
In case (i), we have shown that this system exhibits very rich dynamics, displaying various bifurcations. Fixed-point, oscillatory activity, and seizure-like activity are three activities in this state.
In case (ii), different types of synchronization are observed depending on the internal and external inputs. We have examined in-dept several specific characteristics, such as the time series of each unit and the correlation matrix between them, the mean signals, and finally, their phase portraits of the network. Similar to case (i), steady-state, oscillatory, and seizure-like activities were observed in this state. Moreover, we showed some interesting patterns that were formed at the boundaries between the areas. Afterward, the network behavior in the frequency domain was examined. Delta, theta, and alpha rhythms appeared in our data.\\
Moreover, in case (i), we find that there is no phase transition despite observed behavior changes. It is surprising to find that both the first (discontinuous) and second (continuous) phase transitions result from relaxing the mean-field assumption (case (ii)). While mean-field approximation is a valid and valuable method in a complex systems investigation, it may result in omitting some noteworthy phenomena. Moreover,  this theory is not efficient and accurate in asynchrony cases.\\
We investigated how inputs changes can result in pathological oscillations similar to those observed in epilepsy. Moreover, our findings indicated that delta waves can be generated by altering inputs in a network of Jansen-Rit neural mass models, which has not been previously observed in a single Jansen-Rit neural mass analysis.\\
To put it briefly, a network can have different complex behaviors depending on the randomness of the input, the initial conditions, the noise, and the coupling strength. Although delays between neural populations may lead to intriguing dynamics in networks, we assumed that they were zero in this study. If this were to be taken into account, it could have significant implications.

\section*{Declaration of interests}
The authors declare that they have no known competing financial interests or personal relationships that could have appeared to influence the work reported in this paper.

\section*{Credit Author Statement}
\textbf{Yousef Jamali}: Supervision, Conceptualization, Formal analysis, Investigation, Methodology, Project administration, Resources,  Software, Validation, Writing – review and editing\\

\textbf{Sheida Kazemi}: Conceptualization, Data curation, Formal analysis, Investigation, Methodology, Software, Validation, Visualization, Writing – original draft, Writing – review and editing

\section*{Funding}
This work has been supported in part by a grant from the Cognitive Sciences and Technologies Council with grant No. 8226. In addition, the second author is indebted to the Research Core: ”Bio-Mathematics with computational approach” of Tarbiat Modares University, with Grant No. IG-39706. 








\bibliographystyle{elsarticle-num} 
\bibliography{test}





\end{document}